%% file: main.tex
\documentclass[lettersize,journal]{IEEEtran}
\usepackage{amsmath,amsfonts}
\usepackage{algorithmic}
\usepackage{array}
\usepackage[caption=false,font=normalsize,labelfont=sf,textfont=sf]{subfig}
\usepackage{textcomp}
\usepackage{stfloats}
\usepackage{url}
\usepackage{verbatim}
\usepackage{graphicx}
\def\BibTeX{{\rm B\kern-.05em{\sc i\kern-.025em b}\kern-.08em
    T\kern-.1667em\lower.7ex\hbox{E}\kern-.125emX}}
\usepackage{balance}

\PassOptionsToPackage{table}{xcolor}
\usepackage[utf8]{inputenc} 
\usepackage{hyperref}       
\usepackage{url}            
\usepackage{booktabs}       
\usepackage{amsfonts}       
\usepackage{microtype}      
\usepackage{graphicx}
\usepackage{amsmath}
\usepackage{amsthm}
\usepackage{longtable}
\usepackage{array, makecell}
\usepackage{multicol,xparse,environ}
\usepackage{pifont}
\usepackage{bbm}
\usepackage[raggedrightboxes]{ragged2e}
\usepackage{amsfonts}
\usepackage{algorithmic}

\usepackage{enumitem}
\usepackage{verbatim}

\usepackage[font=bf]{caption}

\usepackage[caption=false,font=normalsize,labelfont=sf,textfont=sf]{subfig}
\usepackage{soul}
\usepackage{multirow,multicol}
\usepackage{balance}
\usepackage{ifthen}
\usepackage{float}
\usepackage{newtxmath} 
\usepackage{bbding}
\usepackage[linesnumbered,ruled,vlined]{algorithm2e}

\usepackage{makecell}

\usepackage{listings}
\usepackage{xcolor}
\usepackage{bibunits}
\defaultbibliographystyle{IEEEtran}

\SetKwInput{KwInput}{Input}                
\SetKwInput{KwOutput}{Output}              

\newcommand{\pa}{\text{MPAA}}
\newcommand{\sa}{\text{IEDA}}
\newcommand{\x}{\boldsymbol{x}}
\newcommand{\e}{\boldsymbol{e}}
\newcommand{\cont}{\boldsymbol{c}}
\newcommand{\txt}{\boldsymbol{r}}
\newcommand{\adapter}{\mathcal{A}_{\text{A}}}
\newcommand{\fsem}{f_{\text{Sem}}}
\newcommand{\D}{\mathcal{A}_{\text{D}}}

\newcommand{\xtgt}{\boldsymbol{x}^{\text{tgt}}}

\newif\ifseparatebib
\separatebibfalse  

\begin{document}

\date{}

\title{Image Prompt Reconstruction Attacks on Distributed MLLM Inference Frameworks}


\author{Xinjian Luo, Hongyan Chang, Jianxin Wei, Yuncheng Wu,\\
Xiaofeng Gao, Meikang Qiu, Ting Yu, and Xue Liu,~\IEEEmembership{Fellow,~IEEE}%
\thanks{\textbf{Xinjian Luo} and \textbf{Xiaofeng Gao} are with Shanghai Jiao Tong University, Shanghai, China. E-mail: \{xinjianluo, gaoxiaofeng\}@sjtu.edu.cn.}%
\thanks{\textbf{Hongyan Chang}, \textbf{Jianxin Wei}, \textbf{Ting Yu}, and \textbf{Xue Liu} are with Mohamed bin Zayed University of Artificial Intelligence, Abu Dhabi, United Arab Emirates. E-mail: changhongyan0530@gmail.com, \{jianxin.wei, ting.yu, steve.liu\}@mbzuai.ac.ae.}%
\thanks{\textbf{Yuncheng Wu} is with Renmin University of China, Beijing, China. E-mail: wuyuncheng@ruc.edu.cn.}%
\thanks{\textbf{Meikang Qiu} is with Augusta University, Augusta, GA, USA. E-mail: qiumeikang@yahoo.com.}%
}

\markboth{Journal of \LaTeX\ Class Files,~Vol.~18, No.~9, September~2020}%
{How to Use the IEEEtran \LaTeX \ Templates}

\maketitle

\begin{abstract}

Distributed large language model (LLM) inference frameworks connect isolated consumer-grade devices for large-scale model inference, substantially reducing hardware constraints. However, recent studies show that intermediate embeddings transmitted among participants can leak private prompts. As LLMs evolve into multimodal LLMs (MLLMs), this risk extends beyond text: image prompts contain rich visual and semantic information, making their intermediate embeddings highly privacy-sensitive. Yet, image-prompt leakage in distributed MLLM inference remains largely unexplored.

In this paper, we investigate privacy risks to input images caused by intermediate embeddings in distributed MLLM frameworks. We first analyze the information flow from image pixels to intermediate representations. Since image and text embeddings are often intertwined across MLLM layers, we design an image embedding extraction algorithm as a prerequisite for reconstruction attacks, achieving $100\%$ extraction accuracy across almost all MLLM layers in our experiments. Building on this, we develop two passive black-box image reconstruction attacks, \pa\ and \sa, reflecting realistic threats from normal participants with limited knowledge and capability. \pa\ performs fine-grained pixel-level reconstruction via patch-wise information extraction and assembly, while \sa\ performs coarse-grained semantic reconstruction through embedding-guided diffusion generation.

We evaluate our attacks on four representative MLLM families: Gemma 3, Phi 4 Multimodal, Qwen 2.5 VL, and Llama 4 Scout. Results show consistently superior reconstruction performance in various settings. We further analyze the effects of MoE architecture, image preprocessing, model size, and text-image dependency on attack performance. To our knowledge, this is the first study of image reconstruction attacks on MLLMs.

\end{abstract}



\begin{IEEEkeywords}
Prompt Reconstruction, MLLM, Distributed Inference.
\end{IEEEkeywords}

\sloppy

\ifseparatebib
\begin{bibunit}
\fi

\input{tex/1-intro}
\input{tex/2-prelim}
\input{tex/3-problem}

\input{tex/4-methods}

\input{tex/5-experiments}

\input{tex/defense}

\input{tex/related-work}

\input{tex/limitation}

\input{tex/conclusion}

\ifseparatebib
\putbib[ref]
\end{bibunit}
\else
\bibliographystyle{IEEEtran}
\balance
\bibliography{ref}
\fi

\clearpage

\ifseparatebib
\begin{bibunit}
\fi

\input{tex/appendix}

\input{tex/ethics}

\ifseparatebib
\putbib[ref]
\end{bibunit}
\fi



\end{document}

%% file: tex/1-intro.tex
\section{Introduction}
\begin{figure*}[t]
\centering
    \includegraphics[width=.9\textwidth]{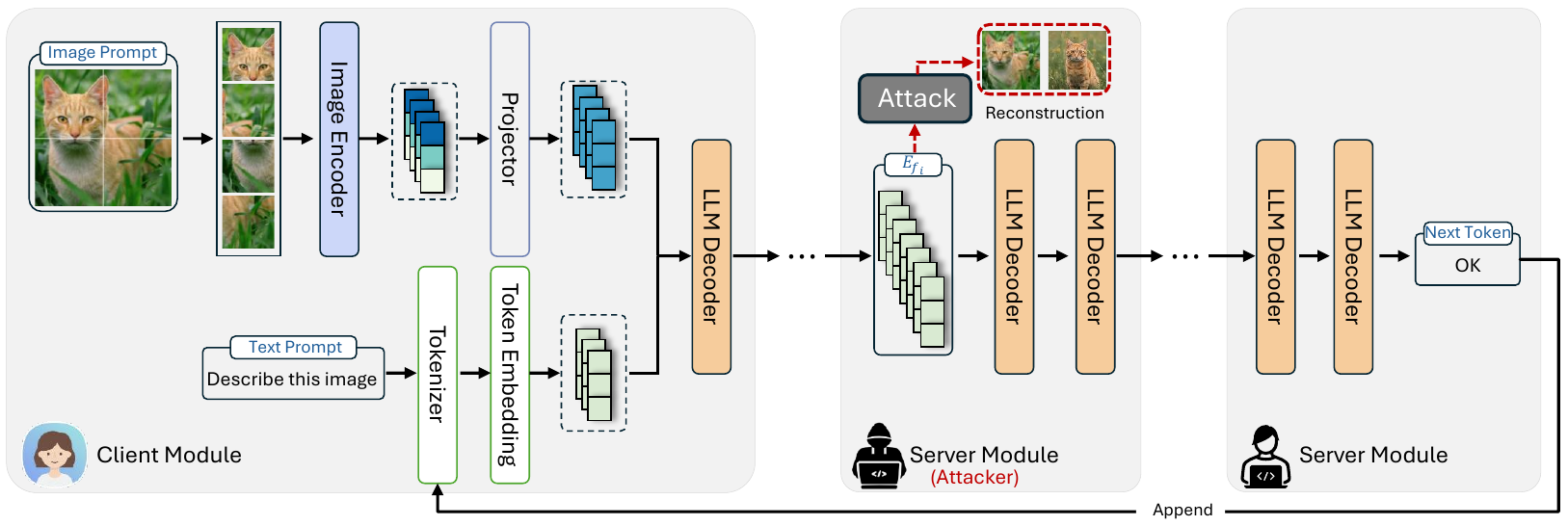}
  \vspace{-3mm}
  \caption{The overview of distributed MLLM inference frameworks.}
  \label{fig-overview}
  \vspace{-3mm}
\end{figure*}

Distributed large language model (LLM) inference frameworks have attracted considerable attention~\cite{Petals-acl,Petals-nips,cake,zhang2024edgeshard-llminference,icml-hexgen-distributedllm,LinguaLinked-distributedllm,sigcomm-pipellm-distributedllm} for their ability to offload large-scale model inference from high-end servers to consumer-grade devices. In parallel, several open-source projects such as Petals~\cite{Petals-acl} and Cake~\cite{cake}, as well as emerging start-ups including Together.ai~\cite{togetherai}, Prime Intellect~\cite{primeintel}, and Modal~\cite{Modal}, have recently built distributed model inference platforms based on this paradigm.
These frameworks typically allocate a subset of LLM layers to each client, enabling multiple clients to collaboratively complete full-model inference using only consumer-grade hardware, as illustrated in Figure~\ref{fig-overview}. 

During the inference process, intermediate representations are transmitted among participants, yet their privacy implications remain insufficiently explored. Although recent studies~\cite{luo2025prompt,qu2025prompt} examine \textit{text prompt} leakage risks in distributed inference, they focus exclusively on text-based LLMs and overlook the privacy threats posed by multimodal LLMs (MLLMs) that process \textit{image inputs}.
This gap raises a fundamental question: \textit{to what extent can private image inputs be inferred from the intermediate representations transmitted along the distributed inference pipeline?} In this work, we address this question by designing image reconstruction attacks targeting distributed MLLM architectures.
To accurately reflect privacy vulnerabilities in real-world frameworks~\cite{Petals-acl}, we assume that any participant in the distributed pipeline may act as an adversary. We further restrict our attack setting to be \textit{black-box} and \textit{passive}. The black-box assumption means that the adversary lacks access to MLLM parameters and architecture details, while the passive assumption indicates that the adversary extracts sensitive information without interfering with the inference process.

It is worth noting that the workflow of distributed MLLM inference shares conceptual similarities with split learning (SL)~\cite{sl,vepakomma2018split}. This naturally raises the question of whether existing attacks on SL~\cite{sl-pcat,sl-tiger,sl-zhundss,erdougan2022unsplit} can be adapted to our setting, as both aim to reconstruct images from intermediate model representations. The answer is negative for two key reasons. First, the target data differs fundamentally: 
SL attacks attempt to reconstruct training images during the model training phase, whereas our attacks aim to recover arbitrary unseen input images during model inference. 
Second, the target models differ significantly: SL attacks operate on lightweight classification networks, where inverse models can be trained efficiently. In contrast, our attacks focus on MLLMs with billions of parameters, for which training inverse models would incur prohibitive resource costs with limited effectiveness.

Another possible direction is to adapt existing text prompt reconstruction attacks on distributed LLMs~\cite{luo2025prompt,qu2025prompt} to reconstruct images in distributed MLLMs. Although these studies demonstrate strong performance on text data, extending them to the image domain faces two fundamental challenges. First, \cite{qu2025prompt} assumes \textit{a white-box LLM setting} in which the adversary needs to host the entire LLM for a successful attack. This assumption violates our black-box and passive threat model and represents \textit{an unrealistic scenario} compared to the typical honest-but-curious participants who have limited resources. Second, \cite{luo2025prompt} exploits \textit{the discrete nature of text tokens} and reconstructs text prompts using classification techniques. However, image pixels are continuous and rich in spatial correlations and fine-grained semantic information, rendering such discrete-space methods \textit{ineffective for visual reconstruction}.
Therefore, developing image reconstruction attacks on MLLMs requires new insights into the information flow within MLLM inference pipelines.

To design simple yet effective attacks, we begin by analyzing the common information flow patterns from input images to intermediate representations across several state-of-the-art MLLMs, including Gemma 3~\cite{gemma3}, Phi 4 Multimodal~\cite{phi4multimodal}, Qwen 2.5 VL~\cite{qwe2.5vl}, and Llama 4 Scout~\cite{llama4}.
Since an MLLM prompt typically contains both text and image data, their corresponding embeddings are intertwined within the intermediate representations produced by different layers.
As a prerequisite, we develop an image embedding extraction method that leverages the distinct properties of text, image, and special tokens in the shared embedding space.
Our experiments show that this method consistently achieves $100\%$ extraction accuracy across all layers in most MLLM architectures.

Based on the extracted image embeddings, we propose two image reconstruction attacks: the \textbf{M}ulti-{r}esolution \textbf{P}atch \textbf{A}ssembly \textbf{A}ttack (\textbf{\pa}) for pixel-level reconstruction and the \textbf{I}mage \textbf{E}mbedding-guided \textbf{D}iffusion \textbf{A}ttack (\textbf{\sa}) for semantic-level reconstruction.
\textit{\pa}\ is inspired by the observation that MLLMs typically split input images into fixed-size patches and generate image embeddings from them, implying that the information of each patch is primarily contained in a subset of embeddings. Accordingly, \pa\ first extracts per-patch pixels from image embeddings and then assembles them into a complete image.
It is important to note that the number of image embeddings does not necessarily match the number of patches, since MLLMs may further split or merge patches during processing, which can cause the loss of visual details.
To enhance reconstruction fidelity, \pa\ employs a multi-resolution reconstruction scheme that first produces two draft images at different resolutions and then fuses them using a fusion-denoising network~\cite{luo2022fusion}.
The high-resolution image is assembled with more pixels per patch, emphasizing fine object details, while the low-resolution image uses fewer pixels per patch to emphasize object structure and contours.
For MLLMs that heavily transform or merge image patches into tokens, especially in deeper layers, the visual information in the output embeddings may be substantially reduced, resulting in degraded performance for \pa.
To address this limitation, we introduce \textit{\sa}, which projects all extracted image embeddings into a predefined semantic space and uses this semantic representation to guide the generation process of a diffusion model.
Compared with \pa, \sa\ focuses on semantic-level reconstruction and demonstrates greater robustness to the loss of fine-grained visual information across different layers of MLLMs.

We conduct extensive experiments on state-of-the-art MLLMs, including Gemma 3~\cite{gemma3}, Phi 4 Multimodal~\cite{phi4multimodal}, Qwen 2.5 VL~\cite{qwe2.5vl}, and Llama 4 Scout~\cite{llama4}, using seven widely adopted datasets: CIFAR10~\cite{CIFAR}, CIFAR100~\cite{CIFAR}, STL10~\cite{STL10}, CelebA~\cite{celeba}, CC3M~\cite{CC3M}, COCO Caption~\cite{COCO}, and ImageNette~\cite{Imagenette}.
Our results show that \pa\ achieves high-fidelity pixel-level reconstruction on the front layers of MLLMs, where visual details remain well preserved within the intermediate embeddings. In contrast, \sa\ achieves consistent semantic-level reconstruction across all layers and all evaluated models.
We further analyze key factors influencing attack performance, including the Mixture-of-Experts (MoE) architecture, image preprocessing strategies, model size, and text-image dependency.
Based on the new findings from these analyses, we discuss potential defense mechanisms against the proposed attacks for the future design of a privacy-preserving distributed inference pipeline.

%

In summary, our main contributions are as follows:
\begin{itemize}
\item We analyze the distributional differences of embeddings corresponding to image, text, and special tokens in MLLMs, and propose an image embedding extraction algorithm as a necessary foundation for image reconstruction attacks. This algorithm consistently achieves nearly $100\%$ extraction accuracy across all layers of most MLLM architectures.
\item We investigate common patterns in MLLM image processing and develop two complementary image reconstruction attacks, \pa\ and \sa, which expose the privacy risks of distributed MLLM inference from both the pixel and semantic levels.
\item We perform a comprehensive evaluation on four representative MLLMs, demonstrating the superior reconstruction capabilities of \pa\ and \sa\ compared to the baselines. Based on the experimental findings, we further explore possible defenses for enhancing the privacy protection of distributed MLLM inference frameworks.
\end{itemize}

%% file: tex/2-prelim.tex
\section{Preliminaries}\label{subsec-MLLM-mechanism}



\noindent
\textbf{Multimodal Large Language Models.}
Mainstream MLLMs are capable of processing both image and text prompts and producing task-dependent outputs, such as text or images.
To jointly handle visual and textual modalities, MLLMs commonly comprise three main components: an \textit{image encoder} $g$, an \textit{image embedding projector}, and a \textit{backbone LLM} $f$. The image encoder converts raw images into embeddings, the projector maps these embeddings into {image tokens}, and the backbone LLM accepts both \textit{image tokens} and \textit{text tokens} as input to generate the final outputs.
Figure~\ref{fig-overview} illustrates a general workflow of this inference process.

Let $\x \in \mathbb{R}^{3 \times H \times W}$ denote an input image with height $H$ and width $W$.
Before being fed into the image encoder, $\x$ is first resized to a predefined resolution, denoted as $3 \times S \times S$, and then divided into patches of a fixed patch size: $X=\{\x_i\}_{i=1}^{N_{\text{patch}}}$, where each patch $\x_i \in \mathbb{R}^{3\times P \times P}$ and $N_{\text{patch}} = \lceil S/P \rceil^2$.
Note that $S$, and hence $N_{\text{patch}}$, may vary depending on the input image size.
The patch set $X$ is subsequently passed through a series of encoder layers to produce $N_{\x}$ patch features.
%
%
It is important to note that typically $N_{\x} \leq N_{\text{patch}}$, as modern MLLMs often employ pooling operations to merge multiple patches, thereby reducing the computational cost of the downstream LLM~\cite{gemma3}.
After obtaining patch features, the embedding projector transforms them into a sequence of image tokens $T^{\x}=\{\boldsymbol{t}_i^{\x}\}_{i=1}^{N_{\x}}$, where each token $\boldsymbol{t}_i^{\x} \in \mathbb{R}^{d_f}$ and $d_f$ denotes the hidden dimension of the LLM $f$.

Let $\txt$ denote the text prompt, and let $T^{\txt}=\{\boldsymbol{t}_i^{\txt}\}_{i=1}^{N_{\txt}}$ represent the text tokens obtained after tokenization and token embedding by the LLM, where $N_{\txt}$ is the sequence length. 
$T^{\txt}$ and $T^{\x}$ will then be concatenated and fed into the LLM for next-token prediction.
Let $L$ denote the number of decoder layers in the LLM $f$, and let $f_i$ represent the $i$-th layer ($i \in {1, \cdots, L}$).
After processing through layer $f_i$, the image and text tokens are transformed into intermediate embeddings $E_{f_i} = \{\boldsymbol{e}_i\}_{i=1}^{N_{\x}+N_{\txt}}$.
%
%
The image information flow in MLLMs can thus be summarized as: $\x\xrightarrow{\text{split}} X \xrightarrow{\text{encoder } \& \text{ projector}} T^{\x}
    \xrightarrow{\text{LLM }f_1 \cdots f_{i}}  E_{f_i}$.


\vspace{0.5mm}
\noindent
\textbf{Distributed MLLM Inference Frameworks.}
To enable MLLM inference on consumer-grade hardware with limited memory, distributed inference frameworks~\cite{Petals-acl,Petals-nips,cake,zhang2024edgeshard-llminference,icml-hexgen-distributedllm,LinguaLinked-distributedllm,sigcomm-pipellm-distributedllm} partition the model into multiple modules according to participants' hardware capabilities. Each participant is then assigned one or more modules to collaboratively perform the full inference process.
We take the widely used Petals framework~\cite{Petals-acl,Petals-nips} as an example.

In Petals, two main modules are defined. A \textit{client module} hosts preprocessing, tokenization, and a small portion of the initial layers (less than $3\%$ of total model weights~\cite{Petals-nips}) and is responsible for initializing inference queries. A \textit{server module} hosts multiple consecutive LLM layers that execute the main inference computation.
Participants may choose to host a client module, a server module, or both~\cite{Petals-nips}.
During inference, intermediate embeddings produced by each front-end server module are transmitted to the next module as its input, enabling a sequential forward pass across participants.
Prior studies~\cite{luo2025prompt,qu2025prompt} have demonstrated that text prompts can be reconstructed from such intermediate text embeddings transmitted between participants. However, the privacy risks associated with image embeddings remain unexplored.

%% file: tex/3-problem.tex
\section{Problem Statement}
\noindent
\textbf{System Model.}
Without loss of generality, we consider a curious participant who hosts a contiguous block of LLM layers beginning at layer $i+1$.  During inference, this participant receives the intermediate embedding $E_{f_i}$ produced by layer $i$ hosted by the preceding participant. When $i=0$, we treat $E_{f_0}$ as the original image tokens input to the LLM, i.e., $T^{\x}$.
When another participant issues a query using a private image $\xtgt$ (the target image), the curious participant can also receive its corresponding embeddings $E_{f_i}^{\text{tgt}}$.

In addition, this participant can also host a client module. Thus, the participant can collect image-embedding pairs prior to layer $i+1$, $\left(\x, E_{f_i}\right)$, using its local client and server components.
Because our attacks are designed to \textit{operate at arbitrary layers}, we omit the layer subscript $f_i$ where convenient and denote intermediate embeddings simply by $E$.

\vspace{0.5mm}
\noindent
\textbf{Threat Model.}
During distributed inference, any participant may act as an adversary. To reflect realistic deployments, we focus on black-box and passive (semi-honest) attacks.
Under the \textit{black-box} assumption, the attacker has no knowledge of remote modules beyond those it hosts. Concretely, the adversary controls a client module $g_{\text{client}}$ and a server module $f_{i+1:j}$ with $j\geq i+1$.
Although this restricts the attacker to a subset of model layers, the black-box model implies two important privacy considerations compared with white-box scenarios~\cite{qu2025prompt} in which an adversary hosts the entire MLLM. 
First, black-box attacks can be executed by ordinary participants using consumer-grade hardware. 
Second, it applies to settings where proprietary MLLMs are hosted within a distributed framework and therefore cannot be inspected in full.
As a result, black-box attacks capture realistic and meaningful privacy risks that white-box analyses may not reveal.
%
%
The \textit{passive} requirement further constrains the adversary: it does not modify or interfere with the inference pipeline.
Instead, the attacker attempts to infer input images solely from the intermediate image embeddings 
$E^{\text{tgt}}$ delivered to its local server module, which makes the attacks stealthy and practical.

Consistent with prior work~\cite{luo2025prompt, sl-pcat,sl-tiger,sl-zhundss}, we assume the attacker can collect an auxiliary image dataset $\mathcal{D}_{\text{aux}}$, which may share domain knowledge~\cite{sl-tiger} or distributional similarity~\cite{luo2025prompt, sl-pcat,sl-zhundss} with the target image $\xtgt$. For example, $\mathcal{D}_{\text{aux}}$ could be a large collection of face images gathered from public sources, while $\xtgt$ is an arbitrary face image.
We emphasize that $\mathcal{D}_{\text{aux}}$ is generally easy to obtain but crucial for attack design.
Unlike tabular or textual data, images exhibit pronounced spatial structure both locally (e.g., correlated neighboring pixels) and globally (e.g., object edges and textures). Effective reconstruction therefore requires learning these spatial patterns from $\mathcal{D}_{\text{aux}}$ so that they can generalize to unseen target images.
In summary, given the received target embedding $E^{\text{tgt}}$, the hosted modules $g_{\text{client}}$ and $f_{i+1:j}$, and the auxiliary dataset $\mathcal{D}_{\text{aux}}$, the attacker seeks to reconstruct the private target image $\xtgt$:
\begin{equation}
    \xtgt \approx \Tilde{\x}^{\text{tgt}} = \mathcal{A}\left(E^{\text{tgt}};g_{\text{client}}, f_{i+1:j}, \mathcal{D}_{\text{aux}}   \right),
\end{equation}
where $\Tilde{\x}^{\text{tgt}}$ denotes the reconstructed image and $\mathcal{A}$ denotes the attack algorithm.

%% file: tex/4-methods.tex
\section{Attack Methods}
In this section, we describe our proposed image reconstruction attacks in detail. As MLLM input prompts comprise both images and text (see Figure~\ref{fig-overview}), a comprehensive prompt-reconstruction attack would ideally recover both modalities. However, a recent study~\cite{luo2025prompt} has already demonstrated high reconstruction accuracy (exceeding $95\%$) for text data under the same attack assumptions. Therefore, we focus exclusively on reconstructing image prompts.

\begin{figure}[!t]
\centering
\begin{minipage}{0.95\linewidth}
\begin{lstlisting}
<bos><start_of_turn>user
You are a helpful assistant.
<start_of_image><image_soft_token>...<image_soft_token><end_of_image>
Describe this image.<end_of_turn>
<start_of_turn>model
\end{lstlisting}
\end{minipage}
\caption{A formatted prompt example for Gemma 3.}
\label{fig-prompt-example}
\vspace{-2mm}
\end{figure}

\begin{figure}[!t]
\centering
\begin{small}
\begin{tabular}{cc}
\subfloat[{In Model Design}]{\includegraphics[width=0.38\columnwidth]{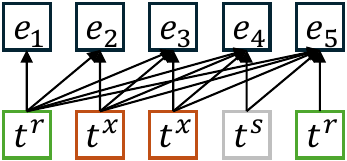}\label{subfig-ar-a}}
&
\hspace{5mm}
\subfloat[{In Practice}]{\includegraphics[width=0.38\columnwidth]{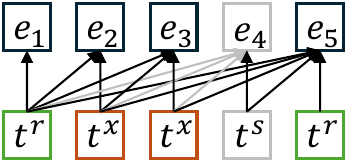}\label{subfig-ar-b}}
\end{tabular}
\caption{The information flow from tokens to embeddings. $t^s$ indicates a special token.}
\label{fig-auto-regressive}
\end{small}
\vspace{-2mm}
\end{figure}

\subsection{Extract Image Embeddings}\label{subsec-extract-embds}
The intermediate embeddings $E$ output by layer $f_i$ contain both image and text embeddings, i.e., $E=E^{\x}\cup E^{\txt}$. To reconstruct the input images, the first step is to extract the image embeddings $E^{\x}$ from $E$. 
An example prompt for Gemma 3 is illustrated in Figure~\ref{fig-prompt-example}. The primary challenge lies in the fact that both $E^{\x}$ and $E^{\txt}$ vary in length. In particular, the number of \texttt{<image\_soft\_token>} depends on the size of the target image, which is unknown to the attacker.

A straightforward approach would be to analyze the distributional differences between image and text embeddings and then train a classifier to distinguish $E^{\x}$ from $E^{\txt}$. 
To examine the feasibility of this approach, we conduct the following experiment. We first collect a set of text prompts from Midjourney Prompts~\cite{midjourney} and use them to generate corresponding images through Midjourney\footnote{\url{https://www.midjourney.com/}}.
This data collection strategy reflects the realistic setting in which image and text components of a multimodal prompt are semantically related. We then feed these image-text pairs into Gemma 3 to obtain their embeddings, apply principal component analysis (PCA)
%
 to reduce their dimensionality to two, and visualize the results in Figure~\ref{fig-motivation-split-image-text} (Supplement~\ref{append-add-results}).
Our analysis reveals that as the embeddings propagate through deeper LLM layers, the representations of image and text tokens become progressively aligned in the embedding space, rendering classification-based separation increasingly ineffective.
Moreover, even in earlier layers, the classifier can hardly achieve perfect accuracy, leading to false-positive image embeddings within  $E^{\x}$ and consequently degrading the reconstruction quality.

\vspace{0.5mm}
\noindent
\textbf{Motivation.}
As the first step of image reconstruction, an ideal embedding extraction method should be analytically simple yet achieve perfect accuracy ($100\%$). To design such a method, we systematically analyze the embedding distributions of different token types across multiple MLLM layers and make a key observation: \textit{embeddings of special tokens (i.e., tokens that carry no semantic input information, such as \texttt{<start\_of\_image>}) are minimally affected by other tokens during inference}.
Note that special tokens are commonly used in modern MLLM designs as structural delimiters.
Because LLMs follow an autoregressive design, each token embedding is influenced by its preceding tokens in the sequence, as illustrated in Figure~\ref{subfig-ar-a}. 
However, we find that the embeddings of special tokens remain nearly constant in the embedding space,
%
%
suggesting that other tokens exert negligible influence on them (Figure~\ref{subfig-ar-b}).
To provide an intuitive analogy, consider a text token such as \texttt{cat}, its embedding can vary depending on the preceding context, e.g., referring to a toy cat or a real cat. In contrast, special tokens act purely as delimiters of user inputs, and their embedding role remains fixed regardless of the surrounding content.

\begin{algorithm}[!tb]
\begin{small}
\caption{Image Embedding Extraction}
\label{alg-image-extraction}
\KwInput{client module $g_{\text{client}}$, server module $f_{i+1:j}$, target embeddings $E^{\text{tgt}}=\{\boldsymbol{e}_i\}_{i=1}^{N}$, minimum gap $\Delta$}
\KwOutput{extracted image embeddings $E^{\x}$}
\BlankLine
\textbf{Notation:} $Q(\cdot;g_{\text{client}}, f_{i+1:j})$ returns server-side embedding sequence\;
Compose a prompt with a random image--text pair and query: $E \!\leftarrow\! Q(\text{prompt}; g_{\text{client}}, f_{i+1:j})$\;
Extract start/end anchors $\boldsymbol{e}_{\text{s}}, \boldsymbol{e}_{\text{e}}$ from $E$\;
Compute distances $d_i^{\text{s}}\!=\!d(\boldsymbol{e}_{\text{s}},\boldsymbol{e}_i)$ and $d_i^{\text{e}}\!=\!d(\boldsymbol{e}_{\text{e}},\boldsymbol{e}_i)$ for $i\!=\!1,\dots,N$\;
Select top-$k$ candidates $\mathcal{I}_{\text{s}}, \mathcal{I}_{\text{e}}$ with smallest $d_i^{\text{s}}$ and $d_i^{\text{e}}$\;
Find $(i_{\text{s}}, i_{\text{e}}) \!=\! \arg\min_{i\in\mathcal{I}_{\text{s}},\, j\in\mathcal{I}_{\text{e}}}\!\! \big(d_i^{\text{s}}\!+\!d_j^{\text{e}}\big)$ 
\textbf{s.t.} $j\!>\!i$ and $j\!-\!i\!\ge\!\Delta$\;
\uIf{$(i_{\text{s}}, i_{\text{e}})$ exists}{
    $E^{\x} \!\leftarrow\! \{\boldsymbol{e}_k\!\in\!E^{\text{tgt}} \mid i_{\text{s}}<k<i_{\text{e}}\}$\;
}
\Else{
    $E^{\x}\!\leftarrow\!\varnothing$ \tcp*{no valid interval}
}
\Return $E^{\x}$\;
\end{small}
\end{algorithm}

\vspace{0.5mm}
\noindent
\textbf{Method.}
Motivated by the stability of special-token embeddings, we design a simple and reliable image embedding extraction procedure as follows. 

\vspace{0.5mm}
\noindent
\textit{Phase 1: Prompt-embedding correspondence.}
The attacker issues a query $\x \cup \txt$ consisting of a random image-text pair to its client module $g_{\text{client}}$ and receives the resulting embeddings $E$ from the server module $f_{i+1:j}$. 
In a realistic deployment, the attacker may observe multiple (say $b$) embedding sequences within the same inference time window, among which only one corresponds to the issued query, while the others are generated by concurrent users.
We can simply establish the correspondence between $\x \cup \txt$ and its embedding $E$ based on sequence length: the number of tokens in $\x \cup \txt$ equals the number of embeddings in $E$. 
Although length collisions are theoretically possible, we observe that they are extremely rare in practice. In particular, experiments on the Midjourney prompt~\cite{midjourney} with $10{,}000$ queries and $b=100$ concurrent embeddings yield a collision rate of $0.2\%$.
To further mitigate this risk, the attacker can optionally insert a short random sequence of special tokens into the text prompt $\txt$ as an identifier. Across all evaluated settings, this combined strategy achieves $100\%$ correspondence accuracy in our experiments.

\vspace{0.5mm}
\noindent
\textit{Phase 2: Image embedding extraction.}
From the prompt structure illustrated in Figure~\ref{fig-prompt-example}, the attacker can directly locate the embeddings corresponding to the special tokens \texttt{<start\_of\_image>} and \texttt{<end\_of\_image>} in $E$. We refer to these vectors as the \textit{anchor embeddings}. 
During the attack process, given a received target embedding sequence $E^{\text{tgt}}$ initialized by other users, the attacker computes pairwise $\ell_2$ or $\ell_1$ distances between each anchor and the vectors in $E^{\text{tgt}}$. 
The nearest vector to an anchor is likely the same special token in the target sequence; and all embeddings that lie between the matched \texttt{<start\_of\_image>} and \texttt{<end\_of\_image>} positions are treated as image embeddings. 
To reduce false positives, we additionally apply top-$k$ candidate selection and a minimum-gap constraint when matching anchors. Further implementation details are given in Algorithm~\ref{alg-image-extraction}.


Let $d_f$ denote the embedding dimension. The pairwise distance computation in Algorithm~\ref{alg-image-extraction} requires $O(|E^{\text{tgt}}|\cdot d_f)$ time, which is substantially more efficient than the classifier-based approach with complexity $O(|E^{\text{tgt}}|\cdot O_{\text{fwd}} )$, where $O_{\text{fwd}}$ represents the forward-pass complexity of the classifier.
In our experiments, Algorithm~\ref{alg-image-extraction} consistently achieves $100\%$ extraction accuracy across all layers of most MLLMs.

\subsection{Pixel-Level Reconstruction (\pa)}
After obtaining the image embeddings $E^{\x} = \{\boldsymbol{e}_i\}_{i=1}^{N_{\x}}$, we train an inverse model $\mathcal{A}_{\pa}$ to recover the input image $\x$ at the pixel level.
In practice, the training dataset $(\mathcal{E}^{\text{aux}}, \mathcal{D}^{\text{aux}})$ for $\mathcal{A}_{\pa}$ can be easily constructed by generating prompts from  $\mathcal{D}^{\text{aux}}$ and forwarding them through the inference pipeline to obtain the corresponding embeddings.
Regarding the design of $\mathcal{A}_{\pa}$, prior studies on image inversion in split learning~\cite{sl-pcat, sl-tiger, sl-zhundss} employ an inverse architecture mirroring the target model $f$. 
While effective for small-scale classifiers with simple convolutional backbones (e.g., ResNet-20~\cite{sl-zhundss}), this approach is impractical for MLLMs containing billions of parameters. Training a full inverse model for such architectures would demand prohibitive resources, contradicting our black-box assumption and the attacker’s limited capability.

\vspace{0.5mm}
\noindent
\textbf{Challenge.}
Due to the autoregressive nature of LLMs (see Figure~\ref{subfig-ar-a}), each embedding may encode information from multiple image patches. A straightforward approach is to concatenate all embeddings into a single vector and use fully connected layers to reconstruct the image as a whole. 
However, this approach faces a critical scalability challenge arising from the high embedding dimensionality of modern MLLMs.
For example, in Phi 4 Multimodal~\cite{phi4multimodal}, an input image of size $128\times 128$ generates $N^{\x}=545$ image embeddings, each with a dimensionality of $d_f=3,072$, resulting in a concatenated vector of $1,674,240$ dimensions. 
Empirically, avoiding overfitting in neural network training requires the dataset size to significantly exceed the input dimension~\cite{nn-memorization}, i.e., $|\mathcal{D}_{\text{aux}}| \gg  1,674,240$.
Collecting a dataset of this scale would demand an impractical number of queries and computational resources, making such a solution infeasible for a resource-limited attacker.
%
%
An effective inverse model should remain lightweight while maintaining generalizability across architectures with diverse embedding sizes.

\begin{figure}[!t]
\vspace{-3mm}
\centering
\begin{small}
\begin{tabular}{cccc}
\hspace{-3mm}
\subfloat[{$\x'$}]{\includegraphics[width=0.22\columnwidth]{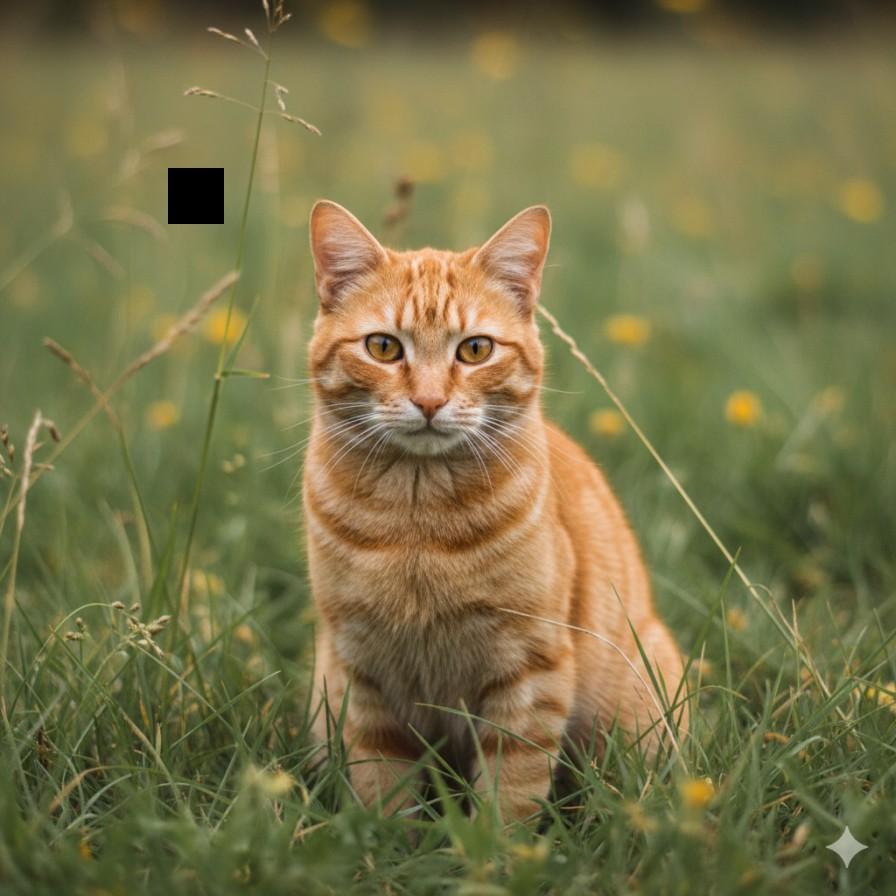}\label{}}
&
\hspace{-2.5mm}
\subfloat[{Layer 4}]{\includegraphics[width=0.22\columnwidth]{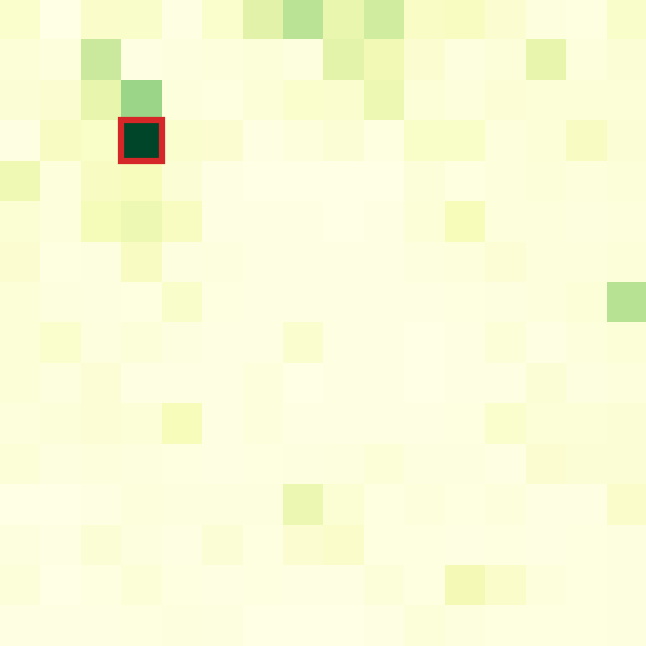}}
&
\hspace{-2.5mm}
\subfloat[{Layer 8}]{\includegraphics[width=0.22\columnwidth]{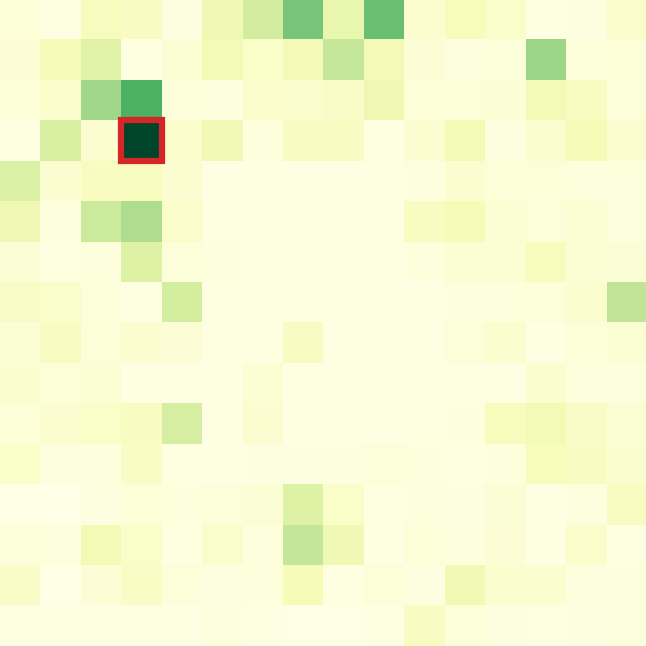}\label{}}
&
\hspace{-2.5mm}
\subfloat[{Layer 16}]{\includegraphics[width=0.22\columnwidth]{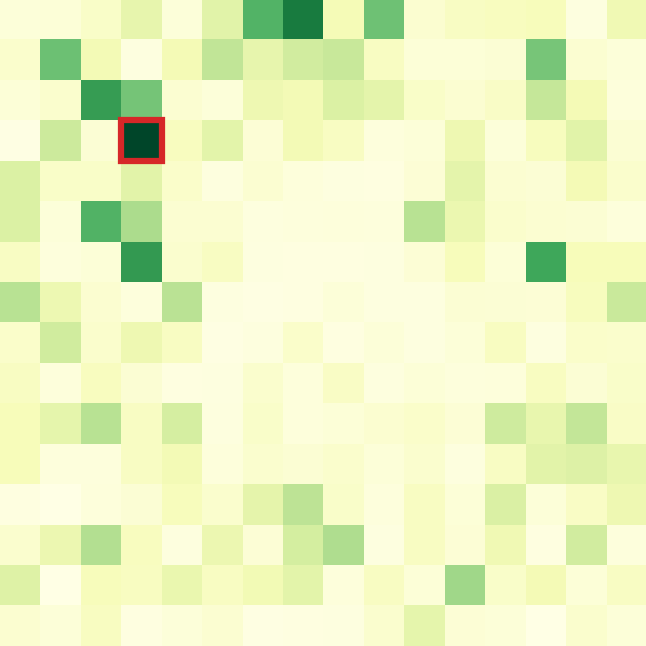}}
\end{tabular}
\caption{Embedding differences caused by masked patches in Gemma 3.}
\label{fig-motivation-patch-impact}
\end{small}
\vspace{-2mm}
\end{figure}

\vspace{0.5mm}
\noindent
\textbf{Motivation.} 
To design a compact inverse model $\mathcal{A}_{\pa}$ with an input dimension much smaller than $N^{\x}\cdot d_f$, we investigate how image information propagates from patches to embeddings.
We conduct the following experiment: an image $\x$ is first fed into Gemma 3 to obtain its layer embeddings $E^{\x}$. We then randomly mask a single patch of $\x$ to form a modified image $\x'$, and feed it into the same model to obtain $E^{\x'}$. 
We compute the $\ell_1$ distance between each pair of embeddings at corresponding positions in $E^{\x}$ and $E^{\x'}$ and visualize the results in Figure~\ref{fig-motivation-patch-impact}, where darker colors indicate stronger impact from the masked patch.
Our key observation is that the masked region primarily affects the embedding at the same spatial position, suggesting that \textit{the information of a given patch is largely contained in its corresponding embedding}.
Therefore, instead of reconstructing the entire image from the joint embedding set, we focus on recovering each image patch from its corresponding embedding independently.

\vspace{0.5mm}
\noindent
\textbf{Method.}
Building on the observation in Figure~\ref{fig-motivation-patch-impact}, we design a pixel-level reconstruction method named the \textbf{M}ulti-{r}esolution \textbf{P}atch \textbf{A}ssembly \textbf{A}ttack (\textbf{\pa}) as follows. 
%

\vspace{0.5mm}
\noindent
\textit{Phase 1: Patch extraction.}
Given the image embeddings $E^{\x} = \{\e_i\}_{i=1}^{N_{\x}}$ with $\e_i\in \mathbb{R}^{d_f}$, we first apply two patch extractors to recover patch-level information at different resolutions:
${\operatorname{Ext}_H}: \e_i \rightarrow \x^H_i$ and
${\operatorname{Ext}_L}: \e_i \rightarrow \x^L_i$,
with $\x^H_i \in \mathbb{R}^{3 \times H \times H}$ and
$\x^L_i \in \mathbb{R}^{3 \times L \times L}$.
Here, $\{H, L\}$ denote the patch resolutions (e.g., $\{16, 8\}$).
We denote the resulting patch sets as $\tilde{X}^H=\{\x^H_i\}_{i=1}^{N_{\x}}$ and $\tilde{X}^L=\{\x^L_i\}_{i=1}^{N_{\x}}$.
A ``patch'' refers to a reconstruction unit that may correspond to one or several adjacent input patches in the client module.
The extracted patches are concatenated to form two complete images, $\tilde{\x}^H$ and $\tilde{\x}^L$, which are then passed through a smoothing network to refine global structure based on spatial continuity between neighboring patches.

\vspace{0.5mm}
\noindent
\textit{Phase 2: Fusion denoising.}
We adapt a Fusion-Denoising Network (FDN)~\cite{luo2022fusion} to reconstruct the input image $\x$ from the two refined images.
Specifically, we first resize the low-resolution image $\tilde{\x}^L$ to size $H \cdot \left\lfloor\sqrt{N_{\x}}\right\rfloor$ (i.e., the size of $\tilde{\x}^H$) using nearest-neighbor interpolation. A convolutional network then extracts the features from both $\tilde{\x}^L$ and $\tilde{\x}^H$, which are further fused using an element-wise averaging rule.
The fused features are decoded into an intermediate reconstruction and subsequently refined by a denoising network to obtain the final output $\tilde{\x}$.
It is important to note that the reconstructed image size $H\cdot\left\lfloor\sqrt{N_{\x}}\right\rfloor$ is not required to match the original input resolution, which is typically unknown to the attacker in realistic scenarios.
That is, \pa\ supports \textit{reconstructing images of arbitrary original sizes to a predefined output resolution}.
The patch extractors, smoother, and FDN are all trained on the pre-collected auxiliary dataset  $(\mathcal{E}^{\text{aux}}, \mathcal{D}^{\text{aux}})$.
%

\vspace{0.5mm}
\noindent
\textit{Remarks}.
The design of \pa\ offers three key advantages: \textit{lightweight}, \textit{adaptive}, and \textit{theoretically grounded}. 
First, \pa\ employs a shared patch extractor for all embedding vectors, which can reduce the model size from $O(N^{\x}d_f)$ in the naive design to $O(d_f)$, leading to significantly lower training data (query) requirements. 
Second, a pre-defined \pa\ can be directly applied to input images of arbitrary sizes and to different MLLM architectures, without requiring structural modification.
Third, our theoretical analysis based on rate-distortion theory and the Shannon lower bound for squared-error distortion~\cite{shannon1959coding} shows that, compared to single-resolution reconstruction, our multi-resolution scheme yields reconstructions that well balance global consistency with local realism.
Detailed discussions are provided in Supplement~\ref{appendix-theoretic-mpaa}.

\begin{figure}[!t]
\centering
    \includegraphics[width=.93\columnwidth]{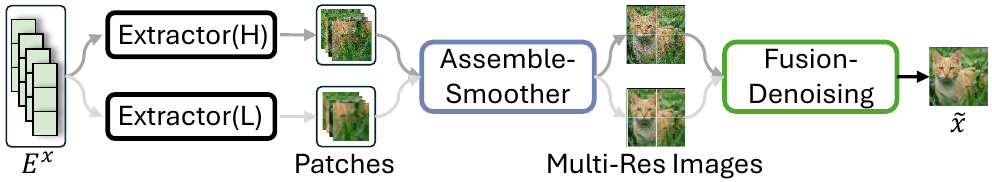}
  \caption{The overview of {M}ulti-{r}esolution {P}atch {A}ssembly {A}ttack (\pa).}
  \label{fig-overview-pa-mpaa}
  \vspace{-3mm}
\end{figure}

\subsection{Semantic-Level Reconstruction (\sa)}
The effectiveness of \pa\ relies on the embedding locality observed in Figure~\ref{fig-motivation-patch-impact}, where the information of neighboring patches is primarily concentrated into a corresponding embedding.
However, Figure~\ref{fig-motivation-patch-impact} also shows that with the processing of deeper layers, the information from each patch gradually diffuses across multiple embeddings. 
%
%
Specifically, embeddings from early LLM layers tend to preserve local visual cues such as edges and textures. As the representations propagate through additional layers, they increasingly capture global semantics such as scene context, becoming more abstract and less spatially detailed. 
Consequently, \pa, which primarily exploits local visual information, exhibits degraded performance at deeper layers.
To reconstruct plausible images across all MLLM layers, we therefore design a semantic-level reconstruction attack that leverages the global semantic information encoded in image embeddings to generate semantically consistent visual content.

\vspace{0.5mm}
\noindent
\textbf{Challenge.}
Unlike \pa, which performs direct pixel-level reconstruction, semantic-level reconstruction is essentially an image generation problem. Among existing generative models, diffusion-based approaches~\cite{DDPM, stable-diffusion}, particularly Stable Diffusion~\cite{stable-diffusion}, are widely recognized for their superior generation quality and sample diversity compared to generative adversarial networks (GANs) or variational autoencoders (VAEs).
Henceforth, a natural strategy is to adapt guided diffusion techniques such as T2I-Adapter~\cite{t2iadapter} and ControlNet~\cite{controlnet}, which control the image generation process using external conditioning signals. These methods typically train an adapter module as a plug-in to a pretrained Stable Diffusion model.
However, we find that such methods are ineffective in our problem. In T2I-Adapter and ControlNet, the external signals represent spatially structured visual features that align naturally with the diffusion model’s generative process, such as depth and sketch~\cite{t2iadapter}. In contrast, our conditional inputs are abstract embeddings produced by MLLMs, which lack explicit spatial organization. As a result, these embeddings are incompatible with the pretrained Stable Diffusion conditioning pathway.
%
%

\vspace{0.5mm}
\noindent
\textbf{Motivation.}
The core rationale behind T2I-Adapter and ControlNet is to align external control images with the internal representations of a pretrained Stable Diffusion model~\cite{t2iadapter}.
We can also utilize this rationale in our semantic-level reconstruction attack. 
Since MLLM embeddings differ fundamentally in modalities from the internal states of diffusion models, we turn to incorporate the embeddings directly as part of the diffusion model’s input to provide generation guidance. 
%
%
The main distinction between our approach and prior methods~\cite{t2iadapter, controlnet} lies in the training method. While those methods keep the diffusion model fixed and train only the adapter, our method jointly trains both the adapter and the diffusion model to achieve tighter alignment between embeddings and the generative process.

\begin{figure}[!t]
\centering
    \includegraphics[width=.9\columnwidth]{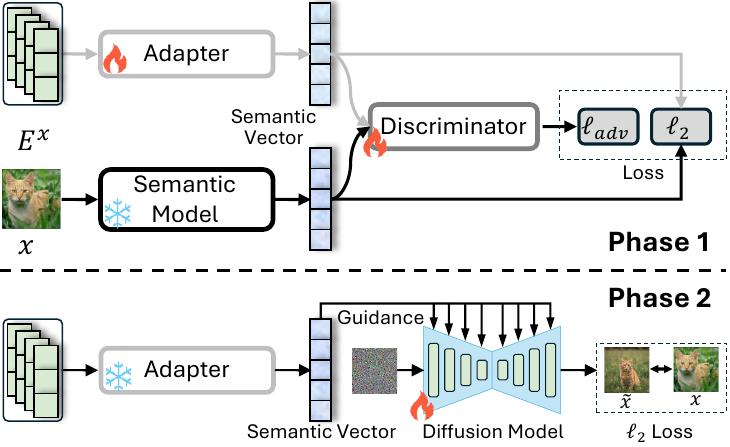}
  \caption{The overview of {I}mage {E}mbedding guided {D}iffusion {A}ttack (\sa).}
  \label{fig-overview-sa-ieda}
  \vspace{-3mm}
\end{figure}

\vspace{0.5mm}
\noindent
\textbf{Method.} 
We design an \textbf{I}mage \textbf{E}mbedding guided \textbf{D}iffusion \textbf{A}ttack (\textbf{\sa}) for semantic-level reconstruction.
As discussed earlier, fixing the diffusion model and training only an adapter can cause signal attenuation, whereas training both the adapter and diffusion model simultaneously may lead to unstable convergence because the control signal generated by the adapter for the same $E^{\x}$ varies continuously during training.
To improve convergence stability while reducing training cost, \sa\ is designed as a two-phase framework. In the first phase, we train only the adapter to project MLLM embeddings into a predefined semantic space. In the second phase, we freeze the adapter and fine-tune a pretrained diffusion model to align the semantic information with the diffusion generation process.
Freezing the adapter helps stabilize convergence, while fine-tuning an existing diffusion model significantly lowers computational overhead compared with training one from scratch. The following sections describe the two training phases of \sa\ in detail.

\vspace{0.5mm}
\noindent
\textit{Phase 1: adapter training}.
The adapter $\adapter$ is designed to produce a control signal $\cont^{\x}$, i.e., a compact semantic vector, from the embeddings $E^{\x}$. We do not directly use $E^{\x}$ as the control input because both the number of embeddings $N^{\x}$ and the embedding dimension $d_f$ vary significantly across different MLLMs. 
To ensure generalization, the architecture of $\adapter$ follows the same design as the patch extractor used in \pa.
To make the adapter output semantically meaningful while maintaining generalization, we align its output space with the semantic feature space extracted by a pretrained semantic model, such as Inception V3~\cite{inceptionv3} or CLIP~\cite{clip}. To achieve this alignment, we design an adversarial training framework that encourages the adapter to generate outputs indistinguishable from those of the semantic model.
Specifically, let $\fsem$ denote the pre-trained semantic model, and  $\D$ denote a discriminator that differentiates between features produced by $\adapter$ and those produced by $\fsem$. We train $\D$ using the following objective:
\[ 
    \mathcal{L}_{\text{adv}} = -\mathbb{E}[\D(\fsem(\x))] + \mathbb{E}[\D(\adapter(E^{\x}))] + \lambda_{\text{gp}}\mathcal{L}_{\text{gp}},
\]
where $\mathcal{L}_{\text{gp}}$ is the gradient panelty term following the design of Wasserstein GAN~\cite{WGAN-GP}:
$
    \mathcal{L}_{\text{gp}} = \mathbb{E}\left[ (||\nabla_{\hat{\cont}} \D( \hat{\cont} )||_2 -1 )^2  \right],
$
with $\hat{\cont}=\alpha \fsem(\x) + (1-\alpha) \adapter(E^{\x})$ and $\alpha \sim U(0, 1)$. The term $\mathcal{L}_{\text{gp}}$ enforces 1-Lipschitz continuity of the discriminator to stabilize the adversarial training.

The adapter is then trained to fool the discriminator while remaining semantically aligned with $\fsem$:
\[
    \mathcal{L}_A=-\mathbb{E} [\D( \adapter (E^{\x}) )  ] +\lambda_{\text{align}}||\adapter (E^{\x}) - \fsem(\x) ||_2^2.
\]
The first term promotes global distribution matching, while the second term enforces element-wise alignment between the semantic features.
After training, the adapter’s output serves as the semantic guidance signal for fine-tuning the diffusion model in the next phase.

\vspace{0.5mm}
\noindent
\textit{Phase 2: diffusion fine-tuning}. 
In this phase, we employ the denoising diffusion probabilistic model (DDPM)~\cite{DDPM} as the diffusion backbone. We do not adopt Stable Diffusion~\cite{stable-diffusion}, as it is primarily designed for text-to-image tasks and thus unsuitable for our reconstruction attack without text inputs.

DDPM training consists of a forward process and a reverse process. The forward process is modeled as a discrete Markov chain~\cite{DDPM}, where input images  $\x^{0}\sim p_{\text{data}}(\x)$ are gradually corrupted by Gaussian noise over $K$ timesteps, producing an image sequence  $\x^{0}\rightarrow \cdots \rightarrow \x^{K}$. Formally,
$
    p(\x^{k} | \x^{k-1})= \mathcal{N}(\x^{k};\sqrt{1-\beta^{k}}\x^{k-1},\beta^{k}\boldsymbol{I}),
$
where $\beta^{k}$ denotes the noise scale at timestep $k\in \{1,\cdots,K \}$.
The final image $\x^{K}$ follows a Gaussian distribution $\mathcal{N}(\boldsymbol{0}, \boldsymbol{I})$, indicating that the input image is completely transformed into noise. 
In the reverse pass, a neural network $\boldsymbol{s}_{\boldsymbol{\theta}}(\x^{k}, k)$ is trained to estimate the score of image density based on $\x^{k}$ for $k\in\{K,\cdots,1\}$.
Starting from Gaussian noise $\x^{K}\sim \mathcal{N}(\boldsymbol{0}, \boldsymbol{I})$ as the input of the reverse pass, DDPM generates new samples iteratively as
$
    \x^{k-1} \leftarrow \frac{1}{\sqrt{1-\beta^{k}}}(\x^{k}+\beta^{k} \boldsymbol{s}_{\boldsymbol{\theta}}(\x^{k}, k)) + \sqrt{\beta^{k}}\boldsymbol{z}^{k},
$
where $\boldsymbol{z}_t \sim \mathcal{N}(\boldsymbol{0}, \boldsymbol{I})$.

To steer the direction of image generation using the semantic vector $\cont^{\x}$, we adapt the classifier-free guidance (CFG) strategy~\cite{CFG}: the control signal $\cont^{\x}$ and timestep $k$ are together input as the guidance during model training:
\begin{equation}\label{eq-loss-sa-ieda}
    \mathcal{L}_{\textbf{\sa}}({\boldsymbol{\theta}}) = \mathbb{E}_{\x^{k}, k, \cont^{\x}}\left[   || \boldsymbol{s}_{\boldsymbol{\theta}}(\x^{k}, k, \cont^{\x}) - \boldsymbol{\epsilon} ||^2_2\right],
\end{equation}
where $\boldsymbol{\epsilon} \sim \mathcal{N}(\boldsymbol{0}, \boldsymbol{I}) $ denotes the injected random noise at timestep $k$ during the forward pass.

Although CFG improves controllability, it can also reduce sampling diversity~\cite{CFG}, which may degrade reconstruction quality in our setting since the semantic signal $\cont^{\x}$ carries far richer information than typical class labels used in \cite{CFG}.
To address this issue, we propose a \textit{weak guidance training strategy} comprising two stages.
In the first stage, we train a base DDPM on ImageNet~\cite{imagenet} with $\cont^{\x}=\boldsymbol{0}$ in Equation~\eqref{eq-loss-sa-ieda}. This enables the diffusion model to learn diverse visual priors.
In the second stage, we fine-tune the base DDPM on the auxiliary dataset $(\mathcal{E}^{\text{aux}}, \mathcal{D}^{\text{aux}})$. In this stage, we follow the strategy of CFG~\cite{CFG} and jointly finetune two score networks, $\boldsymbol{s}_{\boldsymbol{\theta}}(\x^{k}, k)$ and $\boldsymbol{s}_{\boldsymbol{\theta}}(\x^{k}, k, \cont^{\x})$, where the former promotes sample diversity and the latter provides semantic guidance. During reconstruction, their outputs are combined:
\begin{equation}\label{eq-diffusion-reconst}
    \boldsymbol{s}(\x^{k}, k, \cont^{\x}) = (1+\omega)\boldsymbol{s}_{\boldsymbol{\theta}}(\x^{k}, k, \cont^{\x}) - \omega \boldsymbol{s}_{\boldsymbol{\theta}}(\x^{k}, k),
\end{equation}
where $\omega$ denotes the guidance scale controlling the balance between sampling diversity and fidelity.
Note that the base model is trained only once and can be reused across different MLLMs and datasets. Compared with training a diffusion model from scratch, this two-stage strategy substantially reduces computational cost, improves convergence stability, and enhances reconstruction fidelity.

%% file: tex/5-experiments.tex
\section{Experimental Results}\label{sec-experiment}

\begin{table}[!t]
\setlength{\tabcolsep}{4.5pt}
\caption{MLLM details. $L$ denotes layers, and {bold} numbers denote the default setting.}\label{table-mllm}
\centering
\small
\begin{tabular}{cccccc}
\toprule
{MLLM} & {\#params} & $L_g$ & $L_f$ & {Year} & Developer \\
\midrule
Gemma 3 & \textbf{4}/12/27B & 27 & \textbf{34}/48/62  &	2025  & Google\\ 
Phi 4 Multimodal & {6}B & 27 & 32  &	2025  & Microsoft\\ 
Qwen 2.5 VL & 3/\textbf{7}B & 32 & 36/\textbf{28} & 	2025  & Alibaba \\ 
Llama 4 Scout  & 109B & 34 & 48 &	2025  & Meta \\
\bottomrule
\end{tabular}
\vspace{-4mm}
\end{table}

\noindent
\textbf{MLLMs.}
We evaluate our attacks on four series of state-of-the-art MLLMs: \textit{Gemma 3}~\cite{gemma3}, \textit{Phi 4}~\cite{phi4multimodal}, \textit{Qwen 2.5 VL}~\cite{qwe2.5vl}, and \textit{Llama 4 Scout}~\cite{llama4}. 
Since Llama 4 Scout requires approximately 200 GB of GPU memory, which exceeds our available hardware, we offload part of its layers to CPU memory to complete the inference pipeline. The detailed model configurations are summarized in Table~\ref{table-mllm}.
We primarily conduct attacks on the LLM backbone $f$, as results on later layers are backward compatible with those from earlier layers, particularly those in the image encoder  $g$ that precedes $f$.

\vspace{0.5mm}
\noindent
\textbf{Datasets.} 
Five datasets are used in the main text: \textit{CIFAR10}~\cite{CIFAR} (32$\times$32),
\textit{CIFAR100}~\cite{CIFAR} (64$\times$64),
\textit{STL10}~\cite{STL10} (128$\times$128),
\textit{CelebFaces} (CELEBA)~\cite{celeba} (256$\times$256),
and \textit{ImageNet-100}~\cite{imagenet} (160$\times$160).
Note that we deliberately preprocess these datasets into different resolutions to validate the generalization of \pa\ and \sa\ across different input sizes. The first four datasets are used to evaluate attack performance, while ImageNet-100 serves primarily for pretraining the base diffusion model in \sa.
The attack performance evaluated on more realistic datasets, \textit{CC3M}~\cite{CC3M}, \textit{COCO Caption}~\cite{COCO}, and \textit{ImageNette}~\cite{Imagenette}, is reported in Supplement~\ref{append-add-results}.
Each dataset is randomly divided into training, validation, and testing subsets in the ratio of $(64\%, 16\%, 20\%)$.
For evaluation consistency, we set the reconstruction resolution of all images to $64\times64$.
We defer the attack evaluation \textit{w.r.t.} different output resolutions to Supplement~\ref{append-add-results}.
Unless otherwise specified, the default text prompt in each crafted query is \texttt{Describe this image}.


\vspace{0.5mm}
\noindent
\textbf{Evaluation Metric.} We evaluate reconstruction performance using three metrics: mean absolute error ($\ell_1$ loss), Structural Similarity Index Measure (SSIM)~\cite{ssim}, and Cosine Semantic Similarity (CSS)~\cite{luo2025prompt}. The $\ell_1$ loss captures pixel-level accuracy, SSIM measures perceptual similarity between the original image $\x$ and the reconstruction $\tilde{\x}$, and CSS evaluates semantic similarity by computing the CLIP-based cosine similarity between $\x$ and $\tilde{\x}$. Lower $\ell_1$ loss and higher SSIM and CSS indicate better reconstruction quality.

\vspace{0.5mm}
\noindent
\textbf{Baselines.} Existing work on image reconstruction within MLLMs remains limited. To evaluate the performance of our proposed attacks, we adapt two representative image reconstruction methods from split learning as baselines.

\noindent
\textit{PCAT}~\cite{sl-pcat}. PCAT operates by first training a surrogate model to approximate the behavior of the target model, which is not accessible to the attacker. A corresponding decoder is then trained to reconstruct images from the surrogate model’s outputs. PCAT has only been evaluated on classification models such as ResNet~\cite{resnet}. In our experiments, we adapt PCAT to our attack setting by feeding the image embeddings into a feature extractor module, as done in \pa, to reduce the input dimensionality, and then concatenating the reduced features as the decoder’s input.

\noindent
\textit{SDAR}~\cite{sl-pcat}. SDAR is an enhanced version of PCAT that introduces an adversarial loss to improve both surrogate model training and decoder reconstruction. The overall workflow of SDAR closely mirrors that of PCAT. We apply the same adaptations to SDAR as those used for PCAT.

Note that we defer the experimental \textbf{platform} and \textbf{algorithm implementation} to Supplement~\ref{append-add-exp-setting}.

\subsection{Performance of Embedding Extraction}\label{subsec-experiment-embed-extraction}
\begin{table*}[t]
\centering
\caption{Overlap extraction rates of the embedding extraction algorithm. L denotes the last layer.}\label{table-extraction-accuracy}
\setlength{\tabcolsep}{4pt}
\centering
\small
\begin{tabular}{c|c|cccc|cccc|cccc|cccc}
\hline
\multirow{2}{*}{Dataset} &   \multirow{2}{*}{Text Source}  &\multicolumn{4}{c|}{Gemma 3} & \multicolumn{4}{c|}{Phi 4 Multimodal} & \multicolumn{4}{c|}{Qwen 2.5 VL} & \multicolumn{4}{c}{Llama 4 Scout}\\
\cline{3-18}
&  &   9 & 17 & 25 & L (34) &   8 & 16 & 24 & L (32) &   7 & 14 & 21 & L (28) &   12 & 24 & 36 & L (48)  \\
\hline 
\hline
Midjourney & Dependent	& 1.00 &	1.00 &	1.00 &	1.00  & 1.00 & 1.00 & 1.00 &  1.00 & 1.00 & 1.00 & 1.00 &  1.00 & 1.00 & 0.99 & 0.95 & 0.93 \\
CIFAR100 & Independent &	1.00 &	1.00 &	1.00 &	1.00  & 1.00 &	1.00 & 1.00 &  1.00 & 1.00 & 1.00 & 1.00 & 1.00 & 1.00 & 0.99 &	0.95 & 0.92 \\
\hline
\end{tabular}
\vspace{-2mm}
\end{table*}

We first evaluate the performance of our image embedding extraction algorithm (Algorithm~\ref{alg-image-extraction}), which is a critical prerequisite for the subsequent reconstruction stage. If the embeddings extracted by Algorithm~\ref{alg-image-extraction} contain a number of false positives, the accuracy of the downstream reconstruction attacks will be directly degraded.
We utilize the \textit{Overlap Extraction Rate} (OER) as the evaluation metric. Let $E^{\x}$ denote the ground-truth image embeddings and $\tilde{E}^{\x}$ denote the extracted embeddings. OER is defined as $\textbf{OER} = \frac{|E^{\x} \cap \tilde{E}^{\x}|}{|E^{\x}|}$, representing the proportion of ground-truth image embeddings successfully recovered.
To examine the impact of text-image dependence on extraction accuracy, we evaluate two datasets: Midjourney-generated text-image pairs with strong semantic correspondence (Section~\ref{subsec-extract-embds}), and CIFAR-100 images paired with random Midjourney prompts~\cite{midjourney}. Each text-image pair is wrapped into a single prompt and processed through the inference pipeline.
Let $L$ denote the total number of LLM layers. We report OER values for layers $\{\lceil \frac{L}{4}\rceil, \lceil \frac{2L}{4}\rceil, \lceil \frac{3L}{4}\rceil, L\}$ in Table~\ref{table-extraction-accuracy}.

From Table~\ref{table-extraction-accuracy} we draw two main observations.
\textit{First}, the semantic dependence between text and image in the prompt does not affect extraction performance.
\textit{Second}, Algorithm~\ref{alg-image-extraction} achieves a perfect extraction rate ($100\%$) across all evaluated layers for Gemma 3, Phi 4, and Qwen 2.5; however, for Llama 4 Scout, several false positives appear in deeper layers.
This discrepancy is primarily due to differences in embedding-space density. In the first three MLLMs, the special tokens \texttt{<start\_of\_image>} and \texttt{<end\_of\_image>} are sparsely distributed and well separated from other token types. 
In contrast, Llama 4 Scout exhibits tightly clustered embeddings across token types: as embeddings become progressively denser in deeper layers, anchor embeddings may be incorrectly matched using $\ell_1$ distances.
For example, \texttt{<|tile\_x\_separator|>} or \texttt{<|tile\_y\_separator|>} tokens may be mistakenly identified as \texttt{<|image\_end|>} in deeper Llama 4 layers.
Such distribution characteristics may stem from the Mixture-of-Experts (MoE) architecture in Llama 4, where different special tokens can activate similar expert combinations and thus be routed to similar representation subspaces. 
To our knowledge, the embedding-space characteristics of MoE models remain largely unexplored. A deeper investigation, while worthwhile, falls outside the scope of our study, and we defer this to future work.
Importantly, Table~\ref{table-extraction-accuracy} indicates that these errors are relatively limited in practice.

\begin{figure*}[t]
\centering
\begin{small}
\begin{tabular}{cccc}
\multicolumn{4}{c}{\hspace{0mm} \includegraphics[height=4.5mm]{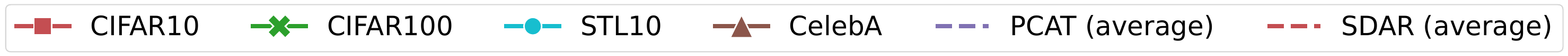}}
\vspace{-4mm}  \\
\hspace{-5mm}
\subfloat[{Gemma 3}]{\includegraphics[width=0.25\textwidth]{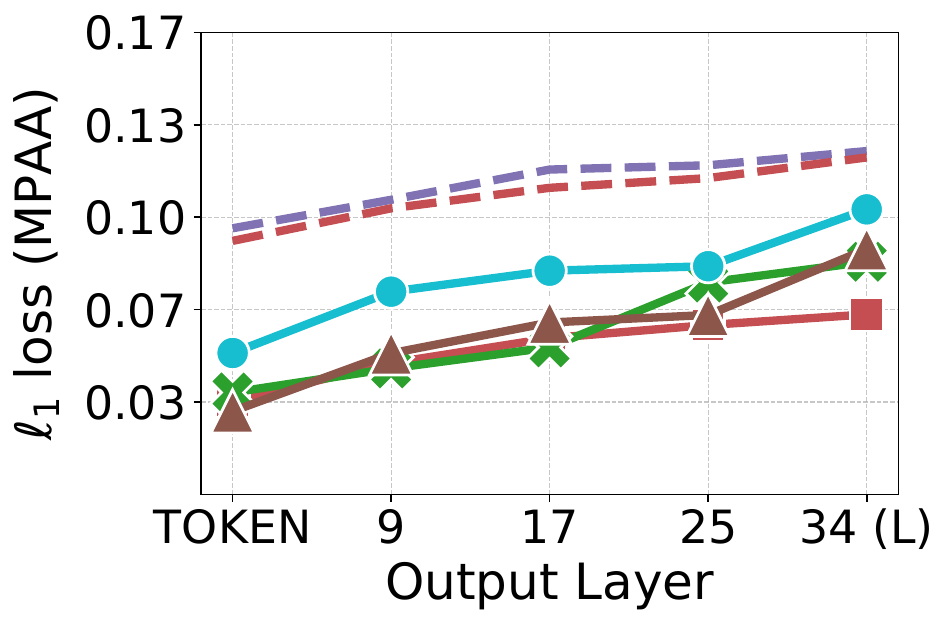}\label{subfig-attack1-l1-l}}
&
\hspace{-5mm}
\subfloat[{Phi 4 Multimodal}]{\includegraphics[width=0.25\linewidth]{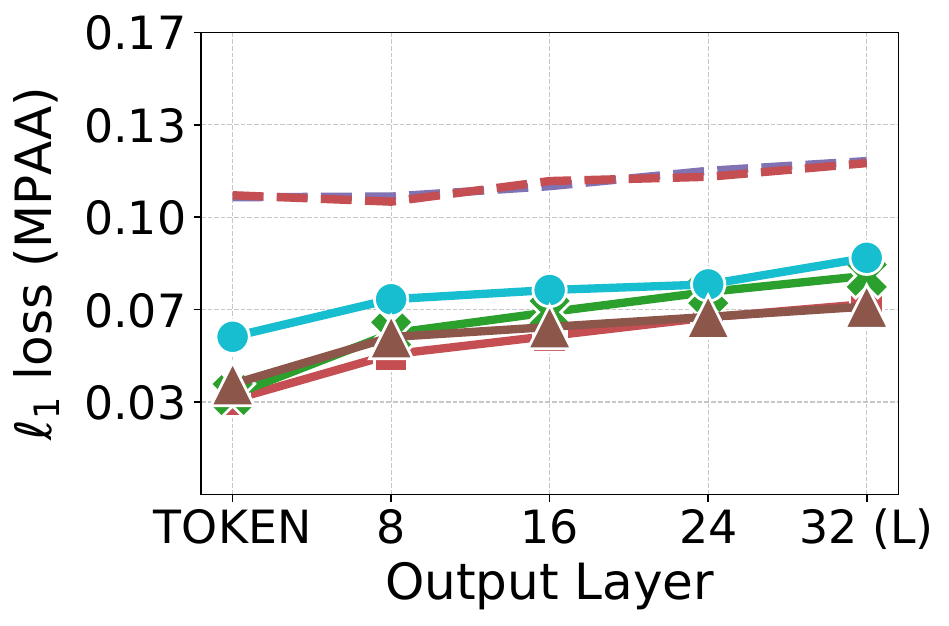}}
&
\hspace{-5mm}
\subfloat[{{Qwen 2.5 VL}}]{\includegraphics[width=0.25\linewidth]{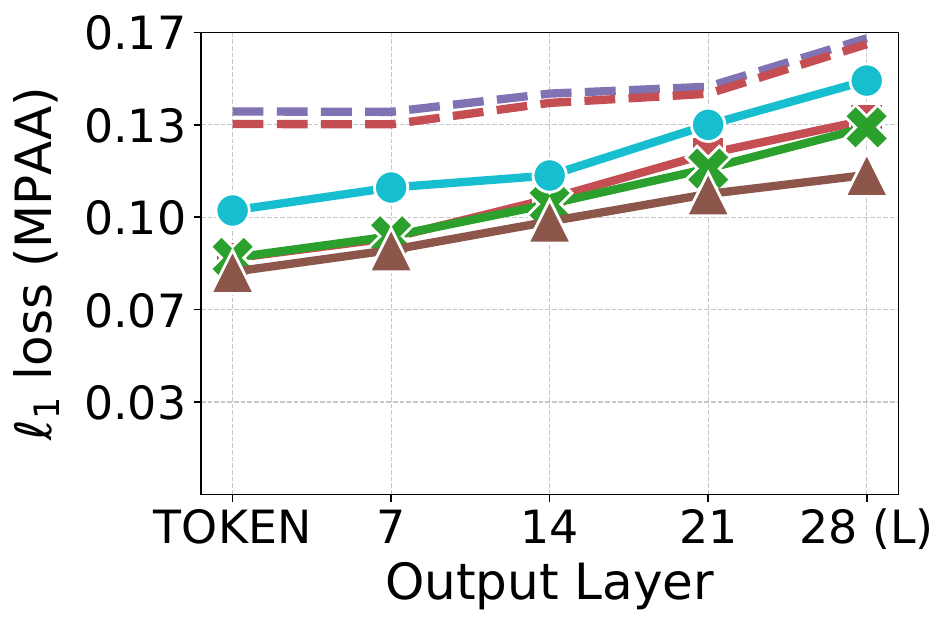}}
&
\hspace{-5mm}
\subfloat[{Llama 4 Scout}]{\includegraphics[width=0.25\textwidth]{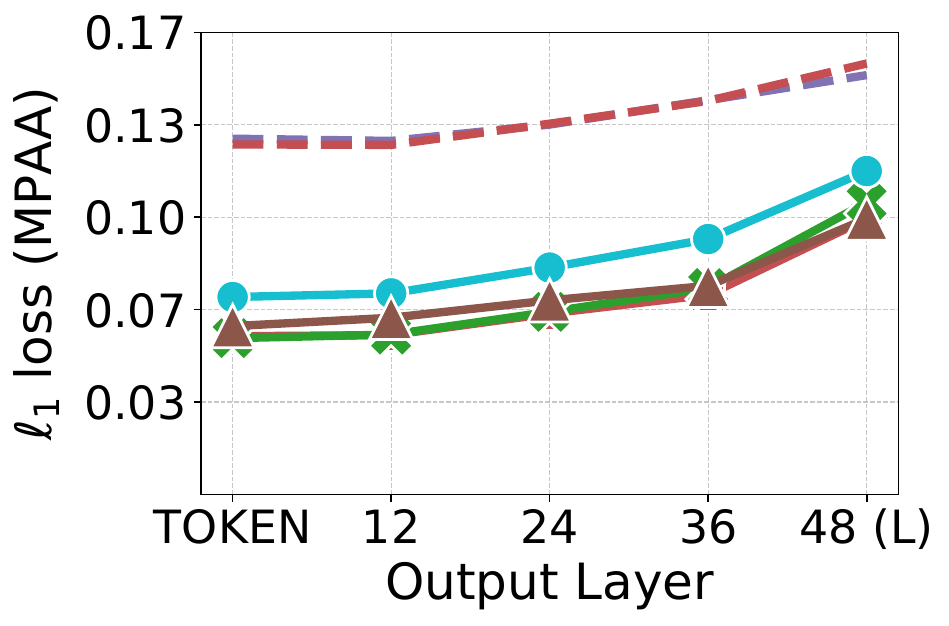}\label{subfig-attack1-l1-r}}
\vspace{-4mm}  \\
\hspace{-5mm}
\subfloat[{Gemma 3}]{\includegraphics[width=0.25\textwidth]{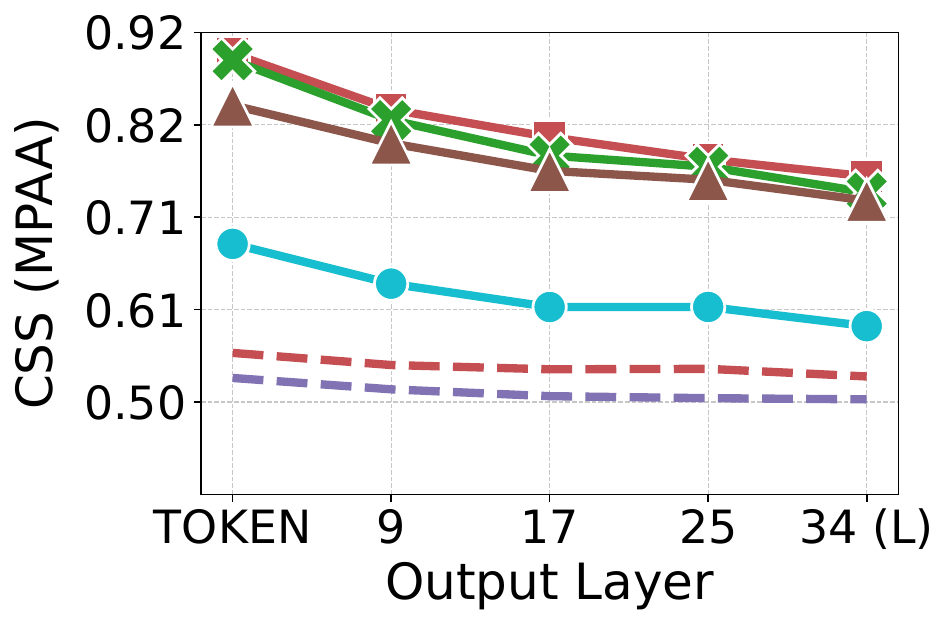}\label{subfig-attack1-css-l}}
&
\hspace{-5mm}
\subfloat[{Phi 4 Multimodal}]{\includegraphics[width=0.25\linewidth]{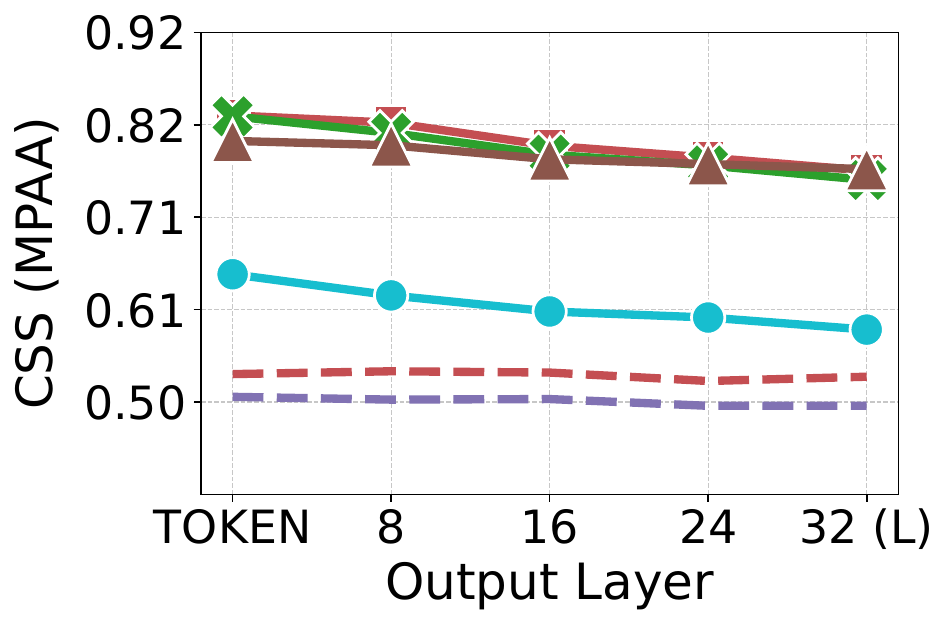}}
&
\hspace{-5mm}
\subfloat[{{Qwen 2.5 VL}}]{\includegraphics[width=0.25\linewidth]{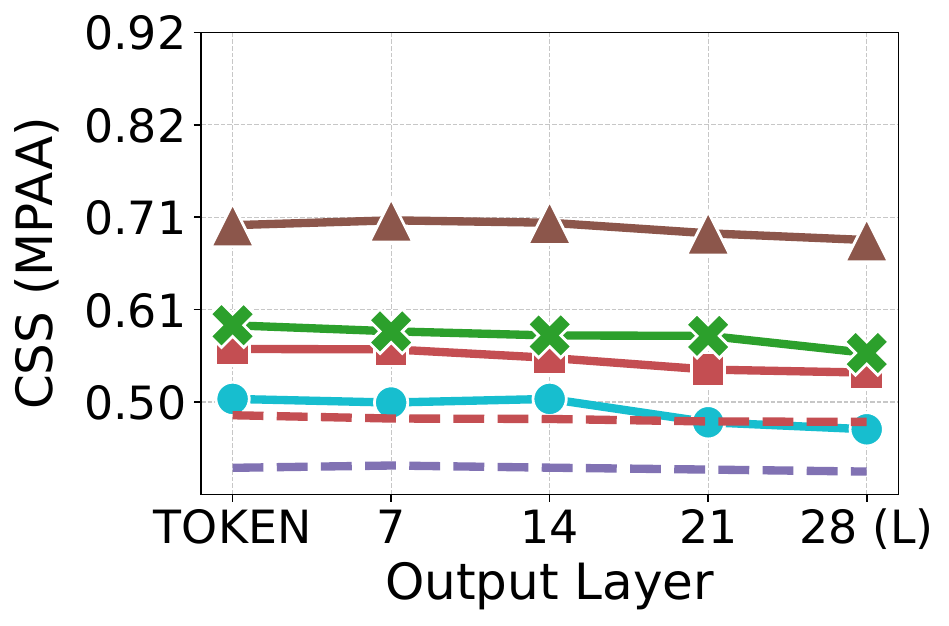}}
&
\hspace{-5mm}
\subfloat[{Llama 4 Scout}]{\includegraphics[width=0.25\textwidth]{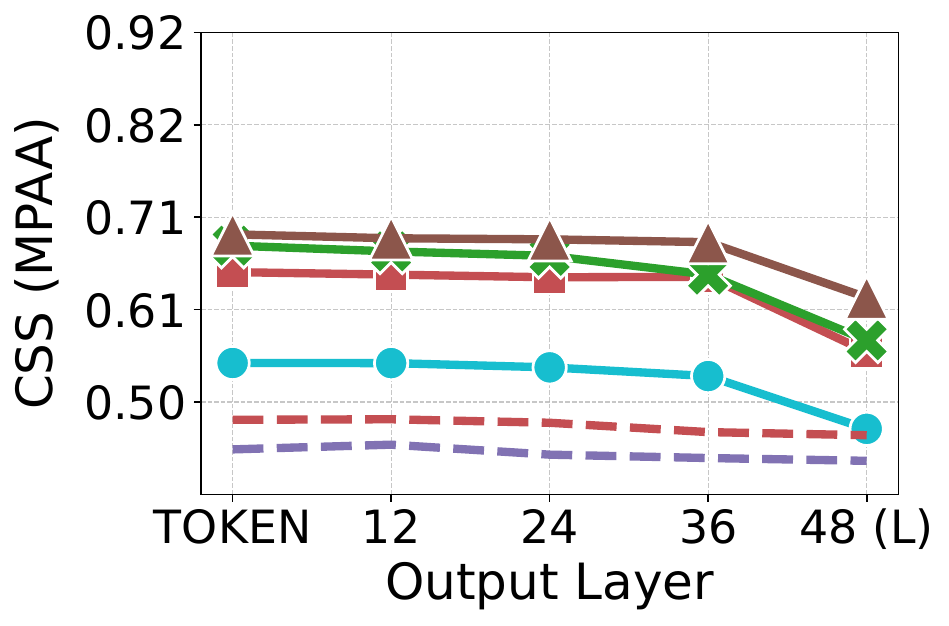}\label{subfig-attack1-css-r}}
\vspace{-4mm}  \\
\hspace{-5mm}
\subfloat[{Gemma 3}]{\includegraphics[width=0.25\textwidth]{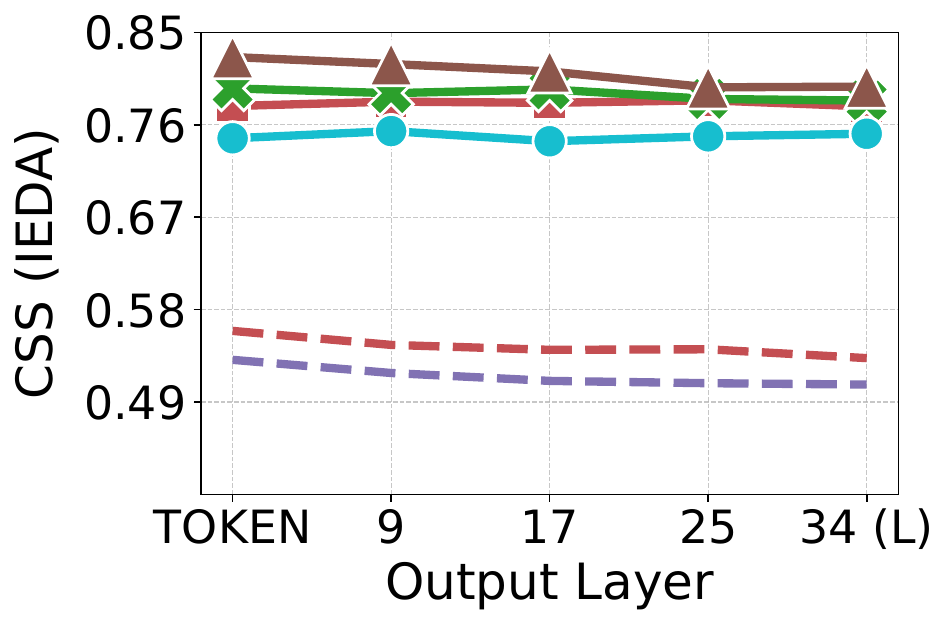}\label{subfig-attack2-css-l}}
&
\hspace{-5mm}
\subfloat[{Phi 4 Multimodal}]{\includegraphics[width=0.25\linewidth]{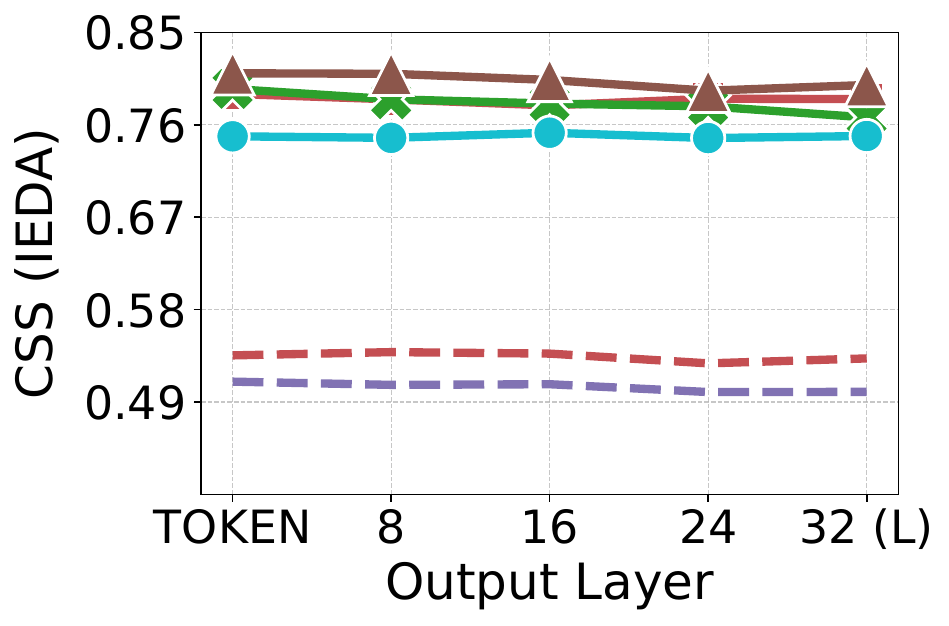}}
&
\hspace{-5mm}
\subfloat[{{Qwen 2.5 VL}}]{\includegraphics[width=0.25\linewidth]{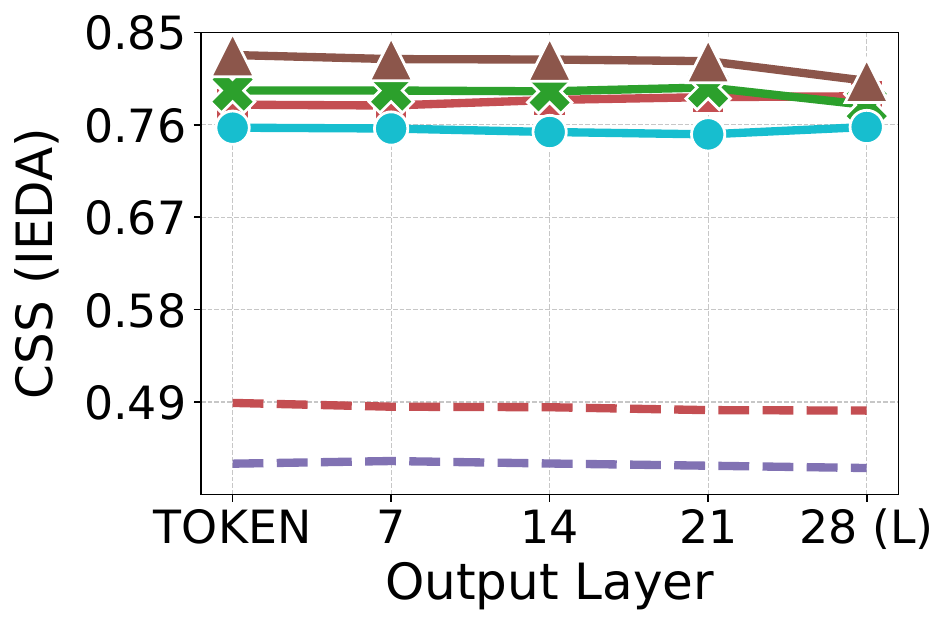}}
&
\hspace{-5mm}
\subfloat[{Llama 4 Scout}]{\includegraphics[width=0.25\textwidth]{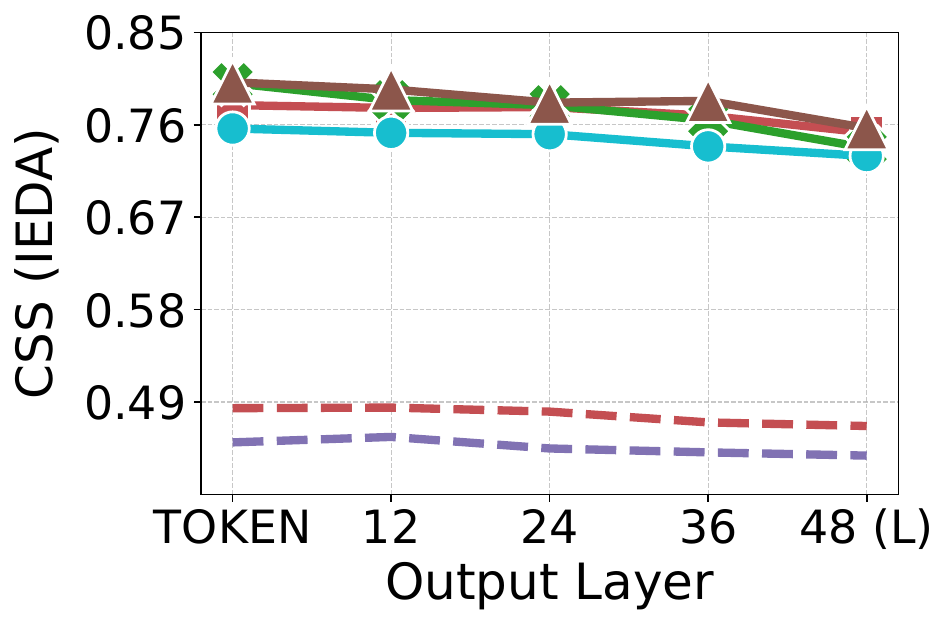}\label{subfig-attack2-css-r}}

\end{tabular}
\caption{
The $\ell_1$ loss ($\downarrow$) of \pa\ ((a)--(b)) and the CSS score ($\uparrow$) of \pa\ ((e)--(h)) and \sa\ ((i)--(l)) across different MLLMs and datasets. Additional results are provided in Figure~\ref{fig-attack-loss-appendix} in the Supplement.
}
\label{fig-attack-loss-main}
\end{small}
\vspace{-2mm}
\end{figure*}

\begin{figure*}
\centering
    \vspace{-3mm}
    \includegraphics[width=.99\textwidth]{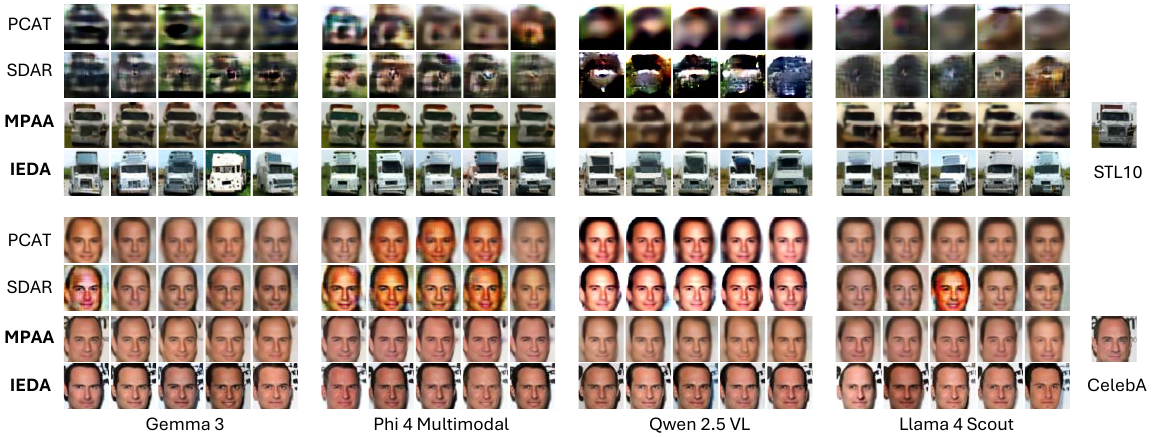}
  \vspace{-1mm}
  \caption{Example images reconstructed by different methods. In each block, the five images from left to right indicate results from the cutting layers $\{0, \lceil \frac{L}{4}\rceil, \lceil \frac{2L}{4}\rceil, \lceil \frac{3L}{4}\rceil, L\}$. See Figure~\ref{fig-attack-examples-appendix} for more examples.}
  \label{fig-attack-examples-main}
  \vspace{-3mm}
\end{figure*}

\subsection{Performance of Image Reconstruction}\label{subsec-experiment-main-result}

We evaluate \pa\ and \sa\ on four datasets across all experimental MLLMs. The cutting layers are set to $\{0, \lceil \frac{L}{4}\rceil, \lceil \frac{2L}{4}\rceil, \lceil \frac{3L}{4}\rceil, L\}$, where $0$ corresponds to the image tokens produced by the image projector (see Figure~\ref{fig-overview}), and $L$ is the total number of layers in the backbone LLM.
We summarize the $\ell_{1}$ loss and CSS of \pa\ in Figure~\ref{subfig-attack1-l1-l}--\ref{subfig-attack1-l1-r} and \ref{subfig-attack1-css-l}--\ref{subfig-attack1-css-r}, respectively. The CSS results for \sa\ are presented in Figure~\ref{subfig-attack2-css-l}--\ref{subfig-attack2-css-r}. 
%
%
Additional reconstruction examples are shown in Figure~\ref{fig-attack-examples-main}.
From Figure~\ref{fig-attack-loss-main} and Figure~\ref{fig-attack-examples-main}, we make three main observations.
\textit{First}, \sa\ consistently outperforms both baselines and \pa\ across all attack settings in terms of semantic similarity. Figure~\ref{fig-attack-examples-main} further illustrates that \sa\ can generate plausible images even at cutting layers where \pa\ performs poorly.
This result is expected because \sa\ leverages both the prior knowledge encoded in the generative model and the semantic information contained in the embeddings.
Since \sa\ does not rely on low-level visual details, the loss of fine-grained visual information in deeper layers does not prevent it from producing semantically meaningful images.
However, the images reconstructed by \sa\ generally preserve only the high-level semantics of the original image while failing to reproduce the precise object structure or spatial arrangement, as shown in Figure~\ref{fig-attack-examples-main}.
\textit{Second}, \pa\ can reconstruct visually plausible images from Llama 4, despite its MoE architecture. This indicates that the MoE structure has limited influence on the preservation of visual information in intermediate embeddings.
Nonetheless, the performance of \pa\ degrades at deeper Llama 4 layers. This degradation is largely due to the false positive cases produced by the image embedding extraction algorithm.
%
%
\textit{Third}, \pa\ performs better on Gemma 3 and Phi 4 than on Qwen 2.5 and Llama 4. We identify two main factors influencing attack performance: the number of image embeddings $N^{\x}$ and the size of the LLM. We provide a detailed discussion of these factors in Section~\ref{subsec-experiment-case-study}.
In addition, both PCAT and SDAR perform worse than the proposed attacks (Figure~\ref{fig-attack-loss-main}), even with the enhancements we apply.
%
%
This result is expected because these methods were originally designed for image classifiers with relatively simple architectures, which are not comparable to modern MLLMs.
Consequently, PCAT and SDAR only produce plausible reconstructions on CelebA (Figure~\ref{fig-attack-examples-main}), while they generally fail on datasets with more complex visual patterns.

In addition, we present the \textbf{attack transferability}, \textbf{ablation studies}, and \textbf{query costs} of our attacks in Supplement~\ref{append-add-results}.

\subsection{Privacy Analysis of MLLMs}\label{subsec-experiment-case-study}
As discussed in Section~\ref{subsec-experiment-main-result}, {\sa\ exhibits stable performance across different settings, whereas \pa\ relies heavily on the visual information preserved} within intermediate embeddings. As a result, \pa\ shows more variable performance across settings.
This observation suggests that \pa\ can \textit{serve as a useful diagnostic tool to assess how well different MLLM components preserve or obscure private visual information}. In what follows, we employ \pa\ to study how various MLLM design choices influence attack performance, from which we derive insights for designing more privacy-preserving MLLMs.

\vspace{0.5mm}
\noindent
\textbf{Text-Image Dependence.} We first examine whether the dependence between text and image in the same prompt affects the attack performance.
Prior work~\cite{mllms-look} shows that textual prompts can influence the attention allocation of MLLMs. For example, a prompt such as \texttt{is there a car in the image} can cause the model to focus more strongly on a car in the background rather than a person in the foreground.
Given these findings, a natural question is whether the choice of textual prompts can affect the performance of \pa.
We begin with an analytical argument: \textit{the design of the textual prompt does not affect the attack performance}.
This follows from two key properties. First, based on the prompt template used in all experimental MLLMs (Figure~\ref{fig-prompt-example}), image tokens always appear before textual tokens. Second, due to the autoregressive nature of LLMs (Figure~\ref{fig-auto-regressive}), only earlier tokens can influence the embeddings of later tokens. Text embeddings that appear after the image tokens therefore cannot modify the image embeddings.
To validate this reasoning, we conduct experiments on Gemma 3 and Phi 4 using CIFAR10 and CelebA.
For each image, we construct two prompts: one with a generic question independent of the image (e.g., \texttt{Describe this image}), and the other with a detailed, image-specific description (e.g., \texttt{a man outdoors, facing right, with his eyes closed or squinting}).
We feed these different image-text pairs into the MLLMs, obtain the corresponding embeddings, and evaluate the attack models used in Figure~\ref{fig-attack-loss-main}.
The $\ell_1$ results are reported in Table~\ref{table-text-image-dependence}.
We observe that attack performance on Phi 4 remains identical across different textual prompts, confirming that its LLM backbone follows the autoregressive property strictly.
A slight fluctuation is observed in Gemma 3. This variation is due to the normalization layers used in Gemma 3, whose internal statistics incorporate all input tokens, causing minor shifts in image embedding distributions. However, as shown in Table~\ref{table-text-image-dependence}, this effect is small and has negligible impact on attack performance.
In conclusion, text-image dependence has minimal influence on the attack performance of \pa.

\begin{table}[t]
\centering
\caption{The $\ell_1$ losses caused by different types of text-image dependence.}\label{table-text-image-dependence}
\setlength{\tabcolsep}{2.6pt}
\centering
\small
\begin{tabular}{c|c|ccc|ccc}
\hline
\multirow{2}{*}{Dataset} &   \multirow{2}{*}{Text Source}  &\multicolumn{3}{c|}{Gemma 3} & \multicolumn{3}{c}{Phi 4 Multimodal} \\
\cline{3-8}
&  &   17 & 25 & L (34) &   16 & 24 & L (32)   \\
\hline 
\hline
\multirow{2}{*}{CIFAR10} & Dependent  &	0.055&	0.061&	0.063 & 0.056&	0.063&	0.068  \\
& Independent	& 0.055&	0.060	&0.064 & 0.056&	0.063	&0.068  \\
\hline
\multirow{2}{*}{CelebA} & Dependent &	0.059&	0.065	&0.080 &	0.059&	0.063	&0.067  \\
& Independent	& 0.061&	0.064	&0.087 &	0.059&	0.063&	0.067  \\
\hline
\end{tabular}
\vspace{-2mm}
\end{table}

\begin{figure}[!t]
\centering
\begin{small}
\begin{tabular}{cc}
\hspace{-5mm}
\subfloat[{STL10}]{\includegraphics[width=0.43\columnwidth]{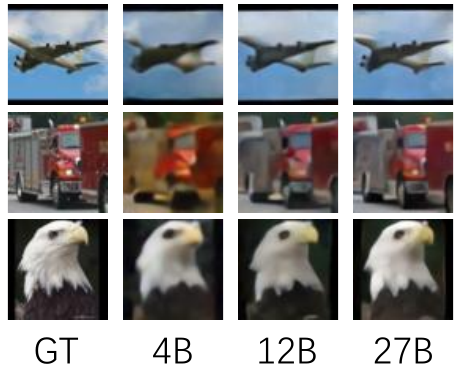}\label{}} &
\hspace{-3mm}
\subfloat[{CelebA}]{\includegraphics[width=0.43\columnwidth]{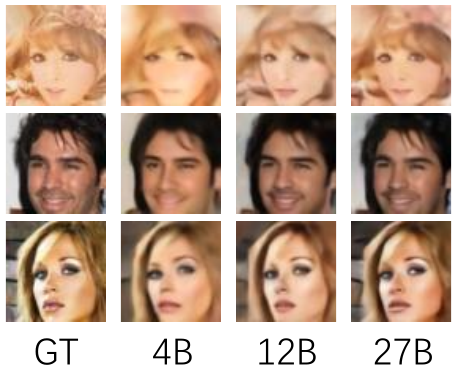}\label{}}  
\end{tabular}
\caption{Reconstructed examples from layer 17 of Gemma 3  under different model sizes.}
\label{fig-diff-model-size-example}
\end{small}
\vspace{-2mm}
\end{figure}

\vspace{0.5mm}
\noindent
\textbf{Model Sizes.} 
The size of the LLM backbone is closely related to both the number of layers and the hidden dimension. As shown earlier in Figure~\ref{fig-attack-loss-main}, processing through deeper layers can degrade the performance of \pa\ to some extent.
Here, we focus specifically on the impact of the LLM \textit{hidden dimension} on attack performance.
For this experiment, we use the Gemma 3 series, which includes three model variants, 4B, 12B, and 27B, with hidden dimensions of 2560, 3840, and 5376, respectively.
All three models share the same image encoder~\cite{gemma3}, which ensures that differences in attack performance are not attributable to variation in the visual encoder.
The $\ell_{1}$ losses on STL10 and CelebA are reported in Table~\ref{table-diff-model-size}, and qualitative examples are shown in Figure~\ref{fig-diff-model-size-example}.
Our main observation is that \textit{\pa\ performs better on MLLMs with larger hidden dimensions}.
The reason is that a larger hidden dimension enables the MLLM to retain more fine-grained visual information in the intermediate embeddings. This richer representation allows the attention mechanism to capture more detailed spatial and semantic cues from the image tokens. Consequently, the embeddings supplied to \pa\ contain more informative visual features, which in turn leads to higher reconstruction quality.

\begin{table}[t]
\centering
\caption{$\ell_1$ losses caused by different Gemma sizes.}\label{table-diff-model-size}
\setlength{\tabcolsep}{7pt}
\centering
\small
\begin{tabular}{c|c|cccc}
\hline
\multirow{2}{*}{Dataset} &   \multirow{2}{*}{Model Size}  &\multicolumn{4}{c}{Layers}  \\
\cline{3-6}
&  & 9 &  17 & 25 & L (34)   \\
\hline 
\hline
\multirow{3}{*}{STL10}  & 4B  &	0.0724	&0.0799	&0.0815&	0.1118  \\
                        & 12B	& 0.0584&	0.0729	&0.0874	&0.0910  \\
                        & 27B	& \textbf{0.0494}&	\textbf{0.0588}	&\textbf{0.0797}	&\textbf{0.0854}  \\
\hline
\multirow{3}{*}{CelebA} & 4B &	0.0504&	0.0614	&0.0641&	0.0876  \\
                        & 12B	& 0.0424&	0.0585&	0.0671&	0.0774  \\
                        & 27B	& \textbf{0.0368}&	\textbf{0.0447}&	\textbf{0.0580}&	\textbf{0.0721}  \\
\hline
\end{tabular}
\vspace{-2mm}
\end{table}

\begin{figure}[!t]
\centering
\begin{small}
\begin{tabular}{cc}
\hspace{-5mm}
\subfloat[{CIFAR10}]{\includegraphics[width=0.45\columnwidth]{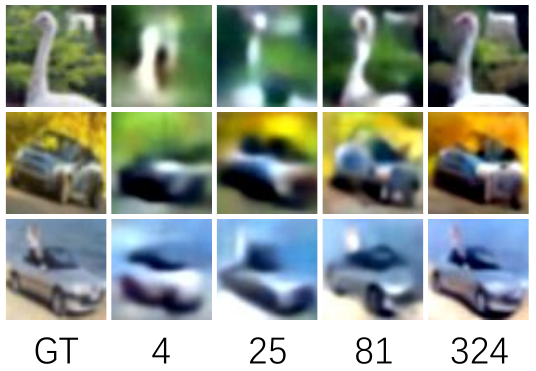}\label{}} &
\hspace{-3mm}
\subfloat[{CelebA}]{\includegraphics[width=0.45\columnwidth]{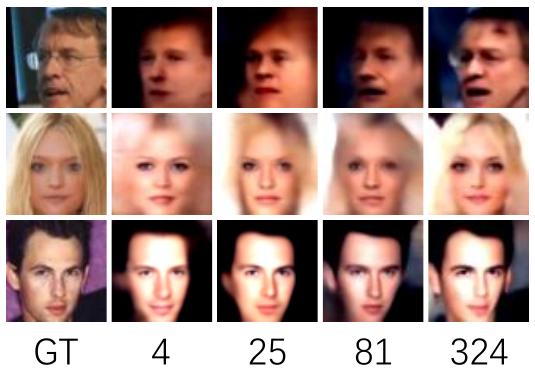}\label{}}  
\end{tabular}
\caption{Reconstructed examples from layer 14 of Qwen 2.5 VL 3B under varying embedding counts.}
\label{fig-diff-patch-example}
\end{small}
\vspace{-2mm}
\end{figure}

\vspace{0.5mm}
\noindent
\textbf{Number of Embeddings.}
We now examine the key factor that explains why \pa\ performs worse on Qwen 2.5 and Llama 4 compared with Gemma 3 and Phi 4, namely, the number of image embeddings. As discussed in Section~\ref{subsec-MLLM-mechanism}, an MLLM first splits the input image into $N_{\text{patch}}$ patches in the preprocessor and extracts $N_{\x}$ image embeddings from these patches.
For the experimental images used in Figure~\ref{fig-attack-loss-main}, the four MLLMs, Gemma 3, Phi 4, Qwen 2.5, and Llama 4, produce $\{256, 256, 4, 144\}$ effective image embeddings, respectively.
This naturally raises the question: Does the extremely small number of embeddings in Qwen 2.5 explain why \pa\ performs the worst on this model?
To answer this question, we conduct the following experiment. We take a dataset with image size 64 and resize each image to different sizes, generating four versions of the same dataset with sizes $\{64, 128, 256, 512\}$.
This ensures that any differences in attack performance reflect the effect of the number of embeddings, rather than differences in image information.
We then train separate attack models on Qwen 2.5 VL 3B using each version.
For image sizes $\{64, 128, 256, 512\}$, Qwen 2.5 extracts $\{4, 25, 81, 324\}$ image embeddings, respectively.
During reconstruction, all images are generated at size 64, since the attacker does not know the original input resolution.
The $\ell_1$ losses for different numbers of embeddings are reported in Table~\ref{table-diff-patch}, and qualitative examples are shown in Figure~\ref{fig-diff-patch-example}.
The observation is straightforward: a larger number of image embeddings leads to greater leakage of private information from the input images.
Compared with the results in Table~\ref{table-diff-model-size} and Figure~\ref{fig-diff-model-size-example}, the effect of the number of embeddings is substantially stronger than that of the LLM hidden dimension.
As illustrated in Figure~\ref{fig-motivation-patch-impact}, each embedding corresponds to one or several specific image patches.
Thus, when an MLLM produces more image embeddings, it effectively captures low-level visual features at a finer granularity, which in turn provides \pa\ with richer visual cues and significantly improves reconstruction performance.
In summary, \textit{the number of embeddings, and therefore the granularity of patch splitting, is the dominant factor underlying the performance differences of \pa\ across MLLMs}.

\begin{table}[t]
\centering
\caption{The $\ell_1$ losses caused by different numbers of embeddings on Qwen 2.5 VL 3B.}\label{table-diff-patch}
\setlength{\tabcolsep}{5pt}
\vspace{-1mm}
\centering
\small
\begin{tabular}{c|c|cccc}
\hline
\multirow{2}{*}{Dataset} &   \multirow{2}{*}{Layer }  &\multicolumn{4}{c}{Image Size (Embedding Num.)}  \\
\cline{3-6}
&  & 64 (4) &  128 (25) & 256 (81) & 512 (324)   \\
\hline 
\hline
\multirow{3}{*}{CIFAR10}  & TOKEN  &	0.1207&	0.1082&	0.0855&	\textbf{0.0658}  \\
                        & 14	& 0.1398&	0.1149&	0.1013&	\textbf{0.0775} \\
                        & 28	& 0.1424&	0.1268&	0.1146&	\textbf{0.0826}  \\
\hline
\multirow{3}{*}{CelebA} & TOKEN &	0.1154&	0.1039&	0.0903&	\textbf{0.0625}  \\
                        & 14	& 0.1200&	0.1057&	0.0985&	\textbf{0.0696}  \\
                        & 28	& 0.1282&	0.1156&	0.1061&	\textbf{0.0773} \\
\hline
\end{tabular}
\vspace{-2mm}
\end{table}

\begin{table}[t]
\centering
\caption{The $\ell_1$ losses of \pa\ and adversarial prediction accuracies under different settings.}\label{table-privacy-robust-trade-off}
\setlength{\tabcolsep}{3.6pt}
\vspace{-1mm}
\centering
\small
\begin{tabular}{c|ccc|cccc}
\hline
\multirow{2}{*}{Metric} &\multicolumn{3}{c|}{Model Size (Gemma)}  &\multicolumn{4}{c}{Embed. Num. (Qwen 3B)}   \\
\cline{2-8}
 & 4B & 12B & 27B & 4 &  25 & 81 & 324   \\
\hline 
\hline
{$\ell_1$ loss ($\downarrow$)} &	0.072&	0.058&	\textbf{0.049} &	0.120&	0.108&	0.085&	\textbf{0.065 } \\
\hline
{APA ($\uparrow$)}  &	44$\%$	& 67$\%$ &	\textbf{69$\%$} &	\textbf{94$\%$} &	87$\%$ &	81$\%$ &	76$\%$  \\
\hline
\end{tabular}
\vspace{-2mm}
\end{table}

\vspace{0.5mm}
\noindent
\textbf{Robustness.} 
The benefits of increasing the hidden dimension and adopting a finer patch-splitting granularity are intuitive: both designs enhance the LLM’s ability to capture subtle visual details, such as a faint traffic sign in the background~\cite{mllms-look}.
However, our earlier results show that these improvements introduce privacy risks in distributed inference.
We now take a further step to examine other potential side effects of improving the model’s capacity for fine-grained visual perception. In particular, we study how these design choices affect the robustness of MLLMs under adversarial perturbations.
To this end, we evaluate different MLLM configurations using the AdvDiffVLM adversarial attack~\cite{AdvDiffVLM}. AdvDiffVLM perturbs a source image belonging to class $A$ such that, when fed into an MLLM, the model outputs a targeted incorrect label (class $B$).
In our experiments, we use \texttt{plane} as the source class and \texttt{car} as the target class (see Figure~\ref{fig-example-adv} for an example). If a perturbed image is still classified as the original class, we consider the prediction robust. Accordingly, we measure robustness via adversarial prediction accuracy (APA), defined as the fraction of perturbed images that remain correctly classified. A higher APA indicates a more robust model.
Table~\ref{table-privacy-robust-trade-off} summarizes the APA of various MLLM configurations, together with the corresponding $\ell_1$ losses of \pa\ for reference.
From Table~\ref{table-privacy-robust-trade-off}, we draw two key findings.
First, model robustness improves as the capability of the LLM backbone increases, as demonstrated by the Gemma 3 series. The 27B model, trained on substantially larger datasets, is better at suppressing irrelevant perturbations and focusing on the true visual features. At the same time, the 27B model is more vulnerable to \pa, illustrating a \textit{privacy-robustness trade-off} when scaling up the LLM backbone.
Second, when the LLM capability is fixed, using fewer embeddings (i.e., adopting a coarser patch-splitting granularity) improves both privacy preservation and robustness. This improvement is likely because coarse patch splitting reduces both the number of embeddings and the amount of low-level visual detail they contain, thereby diminishing the influence of adversarial perturbations.
These observations lead to a practical insight for MLLM architecture design: \textit{when training data and computational resources are limited, using a coarser patch-splitting granularity can simultaneously enhance robustness and privacy; when scaling MLLM size with abundant resources, a fundamental privacy-robustness trade-off emerges}.
We leave a more comprehensive investigation for future work.

%% file: tex/defense.tex
\section{Potential Defenses}

\noindent
\textbf{End-to-End Encryption.}
Since \pa\ and \sa\ rely on plain-text embeddings for image reconstruction, an intuitive defense strategy is to adopt end-to-end encryption. Existing approaches fall into two categories: math-trusted protocols, such as secure multi-party computation~\cite{mp-spdz}, and hardware-trusted sandboxes, such as trusted execution environments (TEEs)~\cite{tee}.
Math-trusted protocols require decomposing non-linear operations (e.g., ReLU) into basic arithmetic operations compatible with encryption schemes. As a result, each non-linear operation requires multiple communication rounds, making inference extremely slow. For example, the encrypted inference system in~\cite{lu2023bumblebee} requires roughly 14 minutes to generate one token on Llama-7B.
Hardware-trusted approaches achieve nearly plain-text efficiency, but require specialized hardware at client side, significantly limiting their practicality for large-scale, real-world distributed MLLM deployments.
In summary, current encryption-based solutions are unlikely to serve as practical defenses for MLLM inference.

\vspace{0.5mm}
\noindent
\textbf{Sophisticated MLLM Design.}
As discussed in Section~\ref{subsec-experiment-case-study}, reducing the number of patches in the preprocessing stage or decreasing the hidden dimension of image embeddings can lower the amount of low-level visual information preserved in the embeddings, thereby weakening \pa.
However, this approach does not meaningfully degrade \sa, which relies primarily on semantic information rather than detailed pixel-level cues.
Moreover, reducing the number of patches diminishes the model’s ability to recognize fine-grained visual details and may harm overall MLLM performance.
Thus, this defense introduces a privacy-utility trade-off that must be carefully managed in practical deployments.

We defer the discussion on \textbf{embedding shuffle}, \textbf{remove special tokens}, and \textbf{differential privacy} to Supplement~\ref{appendix-defense}.

%% file: tex/related-work.tex
\section{Related Work}
\vspace{0.5mm}\noindent
\textbf{Attacks on Language Models.}
A broad range of attacks has been developed to analyze vulnerabilities in large language models, including membership inference~\cite{SongR20promptinference,chang-etal-2025-context}, model extraction~\cite{KrishnaTPPI20modelextraction2,zanella2021greymodelextraction1}, and inference prompt reconstruction~\cite{zhang2024effective-systemprompt,qu2025prompt,luo2025prompt,li2023sentencepromptinference2,MorrisKSR23promptinference,SongR20promptinference,output2prompt}.
%
\textit{Membership inference}~\cite{SongR20promptinference} attempts to determine whether a particular data point was used for model training.
\textit{Model extraction}~\cite{KrishnaTPPI20modelextraction2,zanella2021greymodelextraction1} aims to replicate proprietary fine-tuned models using public LLMs.
\textit{Prompt reconstruction} is most related to ours and includes system prompt reconstruction~\cite{zhang2024effective-systemprompt} and user prompt reconstruction~\cite{qu2025prompt,luo2025prompt,li2023sentencepromptinference2,MorrisKSR23promptinference,SongR20promptinference,output2prompt}.
System prompt reconstruction focuses on recovering hidden system instructions embedded in proprietary APIs, while user prompt reconstruction attempts to infer user-provided inputs either from intermediate embeddings~\cite{li2023sentencepromptinference2,MorrisKSR23promptinference,SongR20promptinference,qu2025prompt,luo2025prompt} or final LLM outputs~\cite{output2prompt}. 
However, all existing prompt reconstruction studies are limited to text-based inputs. Text data consists of discrete tokens, whereas image data involves continuous-valued pixels, making prior methods~\cite{li2023sentencepromptinference2,MorrisKSR23promptinference,SongR20promptinference,qu2025prompt,luo2025prompt} unsuitable for our setting.

\vspace{0.5mm}\noindent
\textbf{Attacks on Vision Models.}
We next review attacks on vision models, mainly under the split learning (SL) paradigm~\cite{vepakomma2018split,sl}.
For image classifiers, reconstruction attacks~\cite{sl-pcat,sl-tiger,sl-zhundss,erdougan2022unsplit} aim to recover inputs from intermediate representations. FSHA~\cite{sl-tiger} performs active attacks by modifying the training process, while PCAT~\cite{sl-pcat}, Unsplit~\cite{erdougan2022unsplit}, and SDAR~\cite{sl-zhundss} perform passive attacks by learning surrogate models and decoders. These approaches, however, target small-scale classifiers during training, whereas our work focuses on inference-time privacy in large MLLMs.
For image embedding encoders, attacks include adversarial methods~\cite{AdvDiffVLM,adv-vision-hu2025transferable} and reconstruction methods~\cite{rect-vision-chen2025leakyclip,rect-vision-lei2025drag}. Adversarial attacks~\cite{AdvDiffVLM} often use transfer-based optimization, generating perturbation masks from surrogate models and transferring them to larger vision architectures such as Llama and Qwen. Reconstruction attacks like LeakyCLIP~\cite{rect-vision-chen2025leakyclip} or DRAG~\cite{rect-vision-lei2025drag} invert CLIP embeddings. DRAG requires white-box access, limiting its applicability beyond small models.
To the best of our knowledge, our work is the first to study image reconstruction attacks on large MLLMs under black-box, distributed inference settings.

%% file: tex/limitation.tex
\section{Discussion}

\noindent
\textbf{Dataset Transferability.}
As discussed in Supplement~\ref{append-add-results}, the performance of \pa\ and \sa\ may degrade under dataset transfer. We emphasize that this degradation is primarily attributable to limited coverage of the training data rather than deficiencies in the attack design itself. Figure~\ref{fig-attack-transferability} supports this conclusion by showing that the degradation occurs mainly when the training and testing datasets exhibit significantly different object patterns.
In principle, a more universal attack could be obtained by training \pa\ and \sa\ on large-scale, diverse image corpora such as LAION-5B~\cite{laion5b}, which was used to train Stable Diffusion and contains billions of images. However, such an approach would incur substantial computational cost, introducing a clear trade-off between attack generality and resource requirements. In this work, our goal is to characterize privacy leakage risks in distributed MLLM inference, rather than to engineer a single attack that generalizes optimally across all possible image distributions.

\vspace{0.5mm}
\noindent
\textbf{Reconstruction Resolution.}
In our experiments, we set the reconstruction resolution to $64$ due to resource constraints. 
Increasing the output resolution does not necessarily yield additional visual details (see Figure~\ref{fig-example-resolution}), as low-level information is progressively lost during MLLM processing.
Notably, whether reconstructions at a resolution of $64$, potentially with missing fine-grained details, constitute meaningful privacy leakage depends on the privacy definitions of specific scenarios. 
While higher-resolution reconstructions with finer details may further benefit an adversary, exploring such settings does not materially alter the conclusions of this work.
%

%% file: tex/conclusion.tex
\section{Conclusion}

%
In this paper, we study privacy vulnerabilities of MLLMs under distributed inference. We analyze information flow from images to intermediate embeddings and propose two complementary attacks: a pixel-level attack (\pa) and a semantic-level attack (\sa). Experiments show that \pa\ recovers fine-grained details at early layers, while \sa\ consistently reconstructs semantic content across all layers. 
Our analysis of factors affecting attack performance offers insights for the design of more privacy-preserving distributed MLLM inference systems.
%

%% file: tex/appendix.tex


\section*{Supplementary Material}
\setcounter{section}{0}
\renewcommand{\thesection}{S\Roman{section}}
\renewcommand{\thesubsection}{S\Roman{section}.\Alph{subsection}}


\section{Design Remarks on \pa}\label{appendix-theoretic-mpaa}
Below, we provide several remarks on \pa.

\vspace{0.5mm}
\noindent
\textbf{Lightweight}. In \pa, we employ a shared patch extractor for all embedding vectors based on the observation that the relationship between image patches and their corresponding embeddings is generally consistent across positions~\cite{luo2025prompt}. Compared with the naive approach that reconstructs the entire image jointly, \pa\ reduces the model size from $O(N^{\x}d_f)$ to $O(d_f)$. 
The smaller model significantly lowers training data requirements, reducing both the number of queries needed and the likelihood of attack detection.

\vspace{0.5mm}
\noindent
\textbf{Adaptive}. \pa\ focuses on modeling the information flow between embeddings and image patches rather than the internal structure of a specific MLLM. After the network architecture is defined, the same design can be applied to different MLLMs with only minor modifications to the input dimension. Moreover, although the input image size of $\x$ affects the number of image embeddings $N^{\x}$, it does not alter the overall attack workflow since the same patch extractor is applied uniformly to all embeddings.

\vspace{0.5mm}
\noindent
\textbf{Multi-resolution reconstruction}. Each input patch $\x_i$ is reconstructed at two resolutions, $\x_i^H$ and $\x_i^L$.
As shown theoretically below using rate-distortion theory and the Shannon lower bound for squared-error distortion~\cite{shannon1959coding}, for patches $\x_i^H$ and $\x_i^L$ of size $L<H$, the expected per-pixel reconstruction distortion satisfies $D_L<D_H$. Hence, low-resolution patches $\x_i^L$ preserve better structural fidelity, while high-resolution patches $\x_i^H$ capture finer visual details. Fusing $\x_i^L$ and $\x_i^H$ thus yields reconstructions that balance global consistency with local realism.

\vspace{0.5mm}
\noindent
\textbf{Theoretical Analysis}. 
Let the original image patch be $\x_i \in \mathbb{R}^{3\times P \times P}$, where $P$ is the patch size, and $\e_i\in \mathbb{R}^{d_f}$ be the embedding output by an MLLM intermediate layer: 
\[
\e_i=f(\x_i).
\]
We define the downsampled patches of $\x_i$ as $\x_i^s$ with $s\in\{L, H\}$ and $L<H\leq P$. The corresponding reconstructions from $\e_i$ are given by 
\[
\hat{\x}_i^s={\text{Extractor}}(\e_i),
\]
and the average per-pixel mean squared error (MSE) is
\[
D_s=\frac{1}{M_s}\mathbb{E}\left[|| \x_i^s - \hat{\x}_i^s ||^2_2   \right],
\]
where $M_s=3s^2$ denotes the number of pixel values in a patch.

According to the Shannon lower bound (SLB) for continuous sources~\cite{shannon1959coding}, we have the minimum achievable distortion (or MSE) for reconstructing $\x_i^s$ from $\e_i$ as:
\begin{equation}\label{eq-distortion}
    D_s\geq N(\x_i^s)2^{\frac{-2I(\x^s_i;\e_i) }{M_s}},
\end{equation}
where $N(\x_i^s)=\frac{1}{2\pi e}2^{\frac{2h(\x_i^s)}{M_s}}$ is the entropy power~\cite{entropy-power} that quantifies the effective variance of image patches, and $I(\x_i^s;\e_i)$ is the mutual information between the image patch and its embedding.
Note that the traditional SLB theory~\cite{shannon1959coding} is derived for i.i.d. sources; we extend it to image patches by modeling each patch $\x_i^s$ as a finite-dimensional correlated Gaussian vector. Under this extension, the distortion lower bound still follows the same exponential dependence on the information rate. 

We now compare the reconstruction distortions between $D_L$ and $D_H$.
There are two components in Equation~\eqref{eq-distortion}, i.e., $N(\x_i^s)$ and $2^{\frac{-2I(\x^s_i;\e_i) }{M_s}}$.

For the first component, when an image patch is downsampled, its high-frequency components will be removed, reducing the entropy power:
\begin{equation}
    N(\x_i^L)<N(\x_i^H),
\end{equation}
as established in signal compression theory~\cite{signal-processing}.

For the second component, we note that the total information contained in the small patch cannot exceed that of the larger patch, leading to
\begin{equation}
    I(\x_i^L;\e_i) \leq I(\x_i^H;\e_i). 
\end{equation}
Although $I(\x_i^L;\e_i) \leq I(\x_i^H;\e_i)$, the denominator $M_s$ (the number of patch pixels) shrinks much faster, specifically $M_H/M_L=(H/L)^2$.
Generally, the information rate per pixel increases as the patch size decreases, that is
\begin{equation}\label{eq-info-rate-per-pixel}
    \frac{I(\x_i^L;\e_i)}{M_L} > \frac{I(\x_i^H;\e_i)}{M_H}.
\end{equation}
Combining Equation~\eqref{eq-distortion}--\eqref{eq-info-rate-per-pixel} yields:
\begin{equation}\label{eq-dl-dh}
    D_L < D_H.
\end{equation}

Equation~\eqref{eq-dl-dh} indicates that smaller patches will have lower per-pixel reconstruction error, hence better reconstruction quality. Specifically, the reconstructed low-resolution patch $\hat{\x}_i^L$ will preserve structural fidelity such as edges, while the high-resolution patch $\hat{\x}_i^H$ provides more high-frequency details like textures. Fusing the multi-resolution patches will provide a balanced reconstruction quality.

\section{Additional Experimental Setting}\label{append-add-exp-setting}
\vspace{0.5mm}\noindent
\textbf{Platform.}
We implement \pa\ and \sa\ in \textit{PyTorch}.
We build a distributed inference environment following Petals~\cite{Petals-acl,Petals-nips} and host it on a server with an AMD EPYC 9654 processor (192 cores), two NVIDIA A100-SXM4 40GB GPUs, and 750 GB of RAM, under Ubuntu 22.04. In our setup, three participants are distributed across the two local GPUs, with the second participant designated as the attacker. Note that the number of participants does not affect the attack mechanism; the primary factor influencing attack performance is the index of the first layer hosted by the attacker.

\vspace{0.5mm}\noindent
\textbf{Algorithm Implementation.} 
For \pa, the patch extractor is implemented as a four-layer multi-layer perceptron (MLP) with batch normalization and LeakyReLU activations after each layer. The input dimension corresponds to the MLLM embedding dimension $d_f$, and the output dimension matches the number of pixels in an extracted patch. We use the implementation of FDN~\cite{luo2022fusion} as the fusion-denoising network, and we adopt a simple averaging rule during patch fusion.
For \sa, the adapter follows the same architecture as the patch extractor in \pa, except for its output dimension. We use the feature extraction part (before the linear layer) of Inception V3~\cite{inceptionv3} as the semantic model, since the linear layer may compress semantic features and thus degrade reconstruction quality. The diffusion model implementation follows DDPM~\cite{DDPM}, augmented with an additional semantic embedding layer before the input stage.

\section{Additional Experiments}\label{append-add-results}
\begin{figure*}[!th]
\centering
    \includegraphics[width=.95\textwidth]{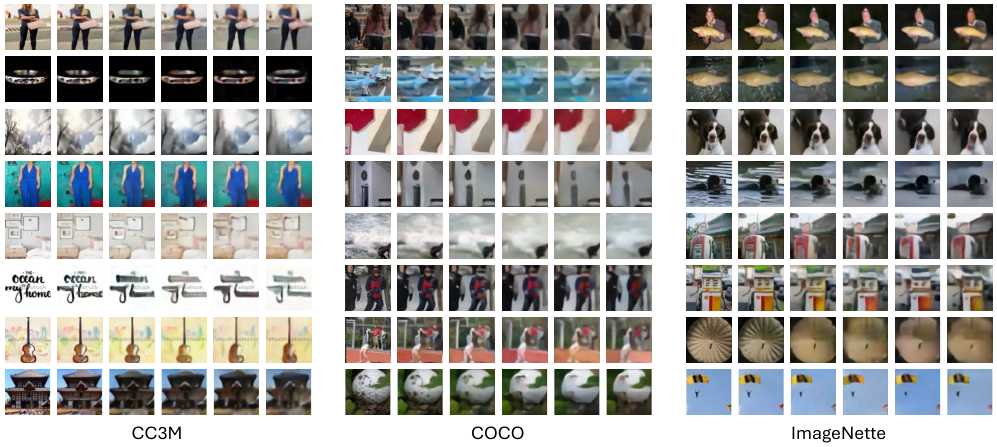}
  \caption{Example images reconstructed via \pa\ from different datasets. In each block, the six images from left to right indicate the ground truth image and results from the cutting layers $\{0, \lceil \frac{L}{4}\rceil, \lceil \frac{2L}{4}\rceil, \lceil \frac{3L}{4}\rceil, L\}$.}
  \label{fig-attack-examples-add_data_pa_appendix}
\end{figure*}

\begin{figure*}[!th]
\centering
    \includegraphics[width=.95\textwidth]{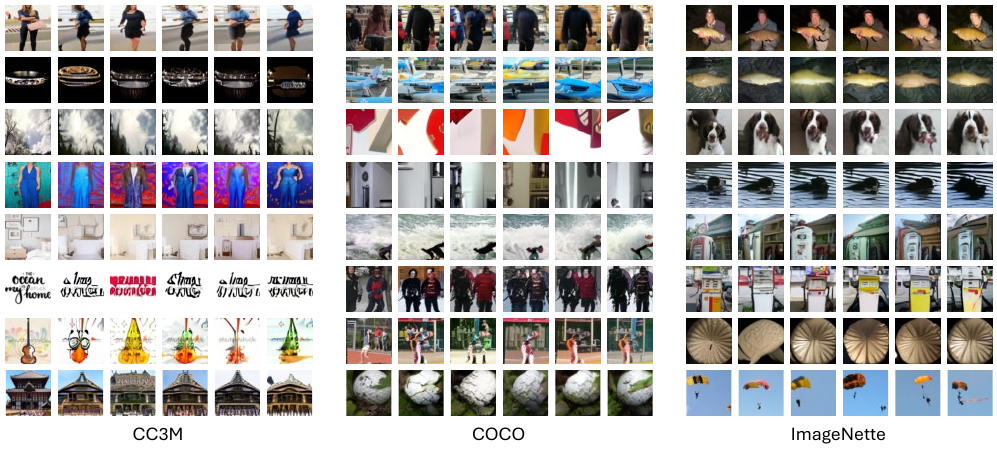}
  \caption{Example images reconstructed via \sa\ from different datasets. In each block, the six images from left to right indicate the ground truth image and results from the cutting layers $\{0, \lceil \frac{L}{4}\rceil, \lceil \frac{2L}{4}\rceil, \lceil \frac{3L}{4}\rceil, L\}$.}
  \label{fig-attack-examples-add_data_sa_appendix}
\end{figure*}

\noindent
\textbf{Real-World Image Datasets.} 
Apart from the classification-based image datasets used in Section~\ref{sec-experiment}, we further evaluate the proposed attacks on three real-world datasets designed for diverse tasks: CC3M~\cite{CC3M}, COCO Caption~\cite{COCO}, and ImageNette~\cite{Imagenette}.
The CC3M dataset contains 3,318,333 image-caption pairs for image captioning tasks and has been used to train multimodal models such as LLaVA~\cite{LLaVA}.
The COCO Caption dataset consists of 328k images with 2.5 million labeled instances and was originally designed to evaluate scene understanding capabilities of MLLMs.
The ImageNette dataset contains 13.4k real-world images and is commonly used as a compact benchmark for transfer learning.
Tables~\ref{table-additional-dataset-tabular-pa} and \ref{table-additional-dataset-tabular-sa} present the attack evaluation of \pa\ and \sa, respectively. The reconstructed example images are shown in Figures~\ref{fig-attack-examples-add_data_pa_appendix} and \ref{fig-attack-examples-add_data_sa_appendix}.
Overall, the observed trends and insights are consistent with those discussed in Section~\ref{subsec-experiment-main-result}.

\begin{table*}[t]
\centering
\caption{The performance of \pa\ evaluated on additional datasets.}\label{table-additional-dataset-tabular-pa}
\setlength{\tabcolsep}{1.8pt}
\centering
\small
\begin{tabular}{c|c|cccc|cccc|cccc|cccc}
\hline
\multirow{2}{*}{Dataset} &   \multirow{2}{*}{Metric}  &\multicolumn{4}{c|}{Gemma 3} & \multicolumn{4}{c|}{Phi 4 Multimodal} & \multicolumn{4}{c|}{Qwen 2.5 VL} & \multicolumn{4}{c}{Llama 4 Scout} \\
\cline{3-18}
&  &  TOKEN & $\lceil \frac{L}{4}\rceil$ &  $\lceil \frac{2L}{4}\rceil$ & $\lceil \frac{3L}{4}\rceil$  &  TOKEN & $\lceil \frac{L}{4}\rceil$ &  $\lceil \frac{2L}{4}\rceil$ & $\lceil \frac{3L}{4}\rceil$&  TOKEN & $\lceil \frac{L}{4}\rceil$ &  $\lceil \frac{2L}{4}\rceil$ & $\lceil \frac{3L}{4}\rceil$  &  TOKEN & $\lceil \frac{L}{4}\rceil$ &  $\lceil \frac{2L}{4}\rceil$ & $\lceil \frac{3L}{4}\rceil$  \\
\hline 
\hline
\multirow{3}{*}{CC3M} & $\ell_1$ loss ($\downarrow$)  &	0.049 &	0.055&	0.063	&0.064  &	0.062&	0.066&	0.070&	0.072 &	0.069&	0.068&	0.076&	0.079 &	0.062&	0.067&	0.070&	0.071 \\
                                & SSIM ($\uparrow$)	&	0.806&	0.767&	0.705&	0.680 &	0.655&	0.613&	0.593&	0.577 &	0.579&	0.582&	0.545&	0.530 &	0.504&	0.524&	0.564&	0.553  \\
                                    & CSS ($\uparrow$) &	0.748&	0.683&	0.641&	0.608  &	0.562&	0.555&	0.549&	0.541 &	0.578&	0.577&	0.550&	0.554 &	0.603&	0.584&	0.589&	0.584\\
\hline
\multirow{3}{*}{COCO} & $\ell_1$ loss ($\downarrow$)  &	0.052&	0.057&	0.061&	0.066  &	0.052&	0.057&	0.061&	0.066 &	0.058&	0.058&	0.062&	0.069 &	0.057&	0.050&	0.066&	0.068 \\
                                    & SSIM ($\uparrow$)	&	0.739&	0.632	&0.555&	0.529 &	0.678&	0.655&	0.521&	0.503 &	0.543 & 0.543&	0.509&	0.478 &	0.596&	0.527&	0.548&	0.522  \\
                                        & CSS ($\uparrow$) &	0.736&	0.685&	0.648&	0.631  &	0.614&	0.614&	0.594&	0.581	&	0.668&	0.667&	0.646&	0.629 &	0.507&	0.520&	0.512&	0.496\\
\hline
\multirow{3}{*}{ImageNette} & $\ell_1$ loss ($\downarrow$)  &	0.047&	0.057&	0.068&	0.067  &	0.063&	0.066	&0.066	&0.067 &	0.069&	0.068&	0.069&	0.075 &	0.068&	0.069&	0.072&	0.070 \\
                                        & SSIM ($\uparrow$)	&	0.807&	0.672&	0.612&	0.603  &	0.578&	0.555&	0.557	&0.540 &	0.534&	0.545&	0.527&	0.521 &	0.679&	0.614&	0.595&	0.612 \\
                                        & CSS ($\uparrow$) &	0.774&	0.718&	0.668&	0.641 &	0.564	&0.551	&0.547	&0.553	 &	0.608&	0.610&	0.597&	0.589 &	0.652&	0.621&	0.551&	0.541\\
\hline
\hline
\end{tabular}
\end{table*}

\begin{table*}[t]
\centering
\caption{The CSS ($\uparrow$) performance of \sa\ evaluated on additional datasets. Note that we omit $\ell_1$ loss and SSIM results for \sa, as these pixel-level metrics cannot accurately evaluate semantic reconstruction quality.}\label{table-additional-dataset-tabular-sa}
\setlength{\tabcolsep}{3pt}
\centering
\small
\begin{tabular}{c|cccc|cccc|cccc|cccc}
\hline
\multirow{2}{*}{Dataset}  &\multicolumn{4}{c|}{Gemma 3} & \multicolumn{4}{c|}{Phi 4 Multimodal} & \multicolumn{4}{c|}{Qwen 2.5 VL} & \multicolumn{4}{c}{Llama 4 Scout} \\
\cline{2-17}
&   TOKEN & $\lceil \frac{L}{4}\rceil$ &  $\lceil \frac{2L}{4}\rceil$ & $\lceil \frac{3L}{4}\rceil$  &  TOKEN & $\lceil \frac{L}{4}\rceil$ &  $\lceil \frac{2L}{4}\rceil$ & $\lceil \frac{3L}{4}\rceil$  &  TOKEN & $\lceil \frac{L}{4}\rceil$ &  $\lceil \frac{2L}{4}\rceil$ & $\lceil \frac{3L}{4}\rceil$ &  TOKEN & $\lceil \frac{L}{4}\rceil$ &  $\lceil \frac{2L}{4}\rceil$ & $\lceil \frac{3L}{4}\rceil$  \\
\hline 
\hline
{CC3M} 	 &	0.835&	0.791&	0.842&	0.826 &	0.724&	0.739	&0.734	&0.734 &	0.738&	0.738&	0.728&	0.749 &	0.732&	0.745&	0.747&	0.735  \\
{COCO}  &	0.799&	0.776&	0.777&	0.700 &	0.781&	0.797&	0.807&	0.767 &0.815&	0.798&	0.717&	0.709&	0.799&	0.758&	0.750&	0.746 \\
{ImageNette} & 0.737&	0.744&	0.753&	0.747 &	0.761&	0.772&	0.738&	0.762 &	0.730&	0.753&	0.776&	0.757 &	0.748&	0.753&0.740	&0.752\\
\hline
\hline
\end{tabular}
\end{table*}

\begin{figure}[!t]
\centering
\begin{small}
\vspace{-5mm}
\begin{tabular}{cc}
\hspace{-5mm}
\subfloat[{TOKEN}]{\includegraphics[width=0.49\columnwidth]{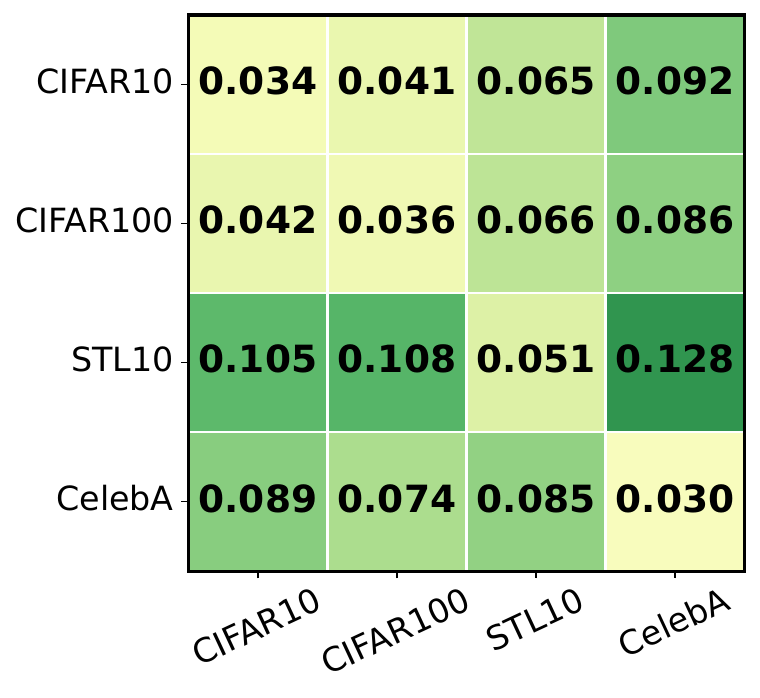}\label{}} &
\hspace{-5mm}
\subfloat[{Layer 9}]{\includegraphics[width=0.49\columnwidth]{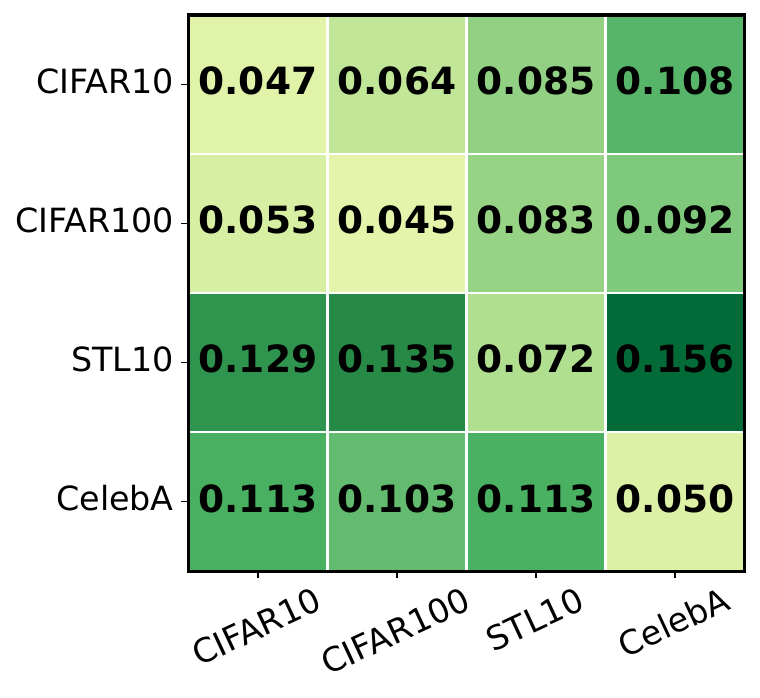}\label{}}  
\end{tabular}
\caption{$\ell_1$ losses for attack transferability by training ($x$-axis) and test ($y$-axis) datasets.}
\label{fig-attack-transferability}
\end{small}
\end{figure}

\vspace{0.5mm}
\noindent
\textbf{Attack Transferability.}
We evaluate the transferability of the proposed attacks, with the goal of answering the following question: Can attack models trained on one dataset be used to reconstruct images from another dataset?
Since \sa\ is a diffusion-based image generation method, its reconstruction quality largely depends on the diversity of its training data. In particular, \sa\ is unlikely to generate objects that do not appear in its training set~\cite{CFG}.
For this reason, our transferability analysis primarily focuses on \pa.
In this experiment, we train attack models on four datasets, CIFAR10, CIFAR100, STL10, and CelebA, using cutting layers $\{0, 9\}$ of Gemma 3. We then evaluate each trained model on the three remaining datasets that differ from its training data.
The $\ell_1$ losses of this transferability study are shown in Figure~\ref{fig-attack-transferability}, from which we draw two main observations.
\textit{First}, transferability is influenced by the distribution similarity between the training and test datasets. For instance, Figure~\ref{fig-attack-transferability} shows that models trained on CIFAR10 and CIFAR100 transfer well to each other, as these datasets share similar visual and semantic distributions. In contrast, models trained on CelebA transfer poorly to the other datasets because human faces differ substantially from the natural objects present in CIFAR10, CIFAR100, and STL10.
\textit{Second}, deeper cutting layers reduce attack transferability. The LLM backbone primarily serves to align text and image embeddings, and as more layers process the image embeddings, their distribution gradually shifts into a different representation space. This shift increases the distribution mismatch between datasets and thus degrades the attack transferability.

\begin{figure}[!t]
\centering
\begin{small}
\begin{tabular}{cc}
\multicolumn{2}{c}{\hspace{0mm} \includegraphics[height=4.5mm]{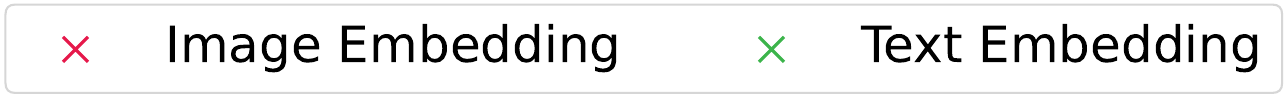}}
\vspace{-4mm}  \\
\hspace{-5mm}
\subfloat[{Layer 8}]{\includegraphics[width=0.4\columnwidth]{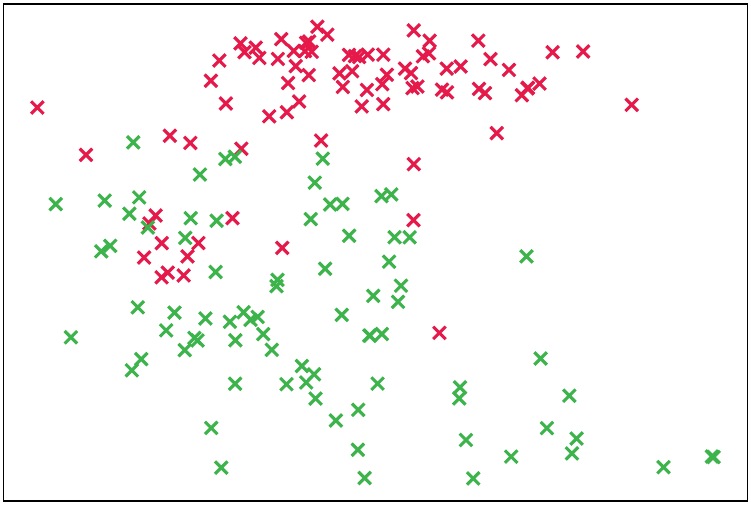}\label{subfig-motivation-l8}}
&
\subfloat[{Layer 33}]{\includegraphics[width=0.4\columnwidth]{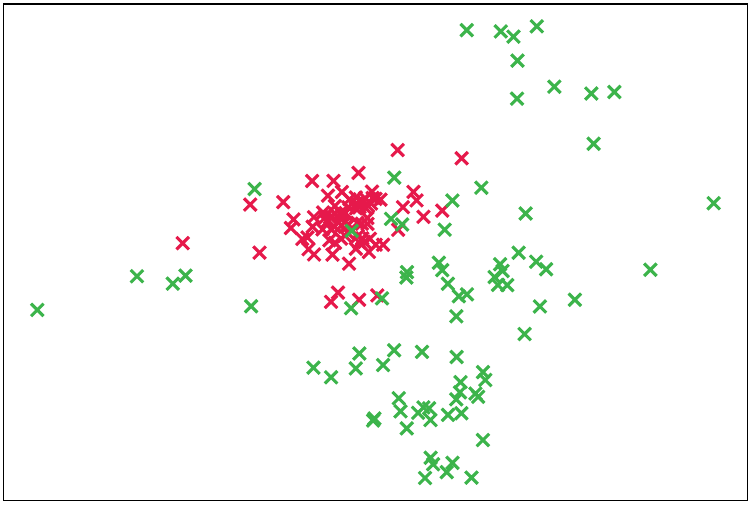}}
\end{tabular}
\caption{Text and image distributions in Gemma 3.}
\label{fig-motivation-split-image-text}
\end{small}
\end{figure}

\begin{figure}[!t]
\centering
\begin{small}
\begin{tabular}{cc}
\subfloat[{\pa}]{\includegraphics[width=0.7\columnwidth]{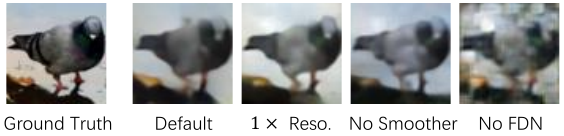}\label{subfig-ablation-attack1-egg}} \\
\subfloat[{\sa}]{\includegraphics[width=0.7\columnwidth]{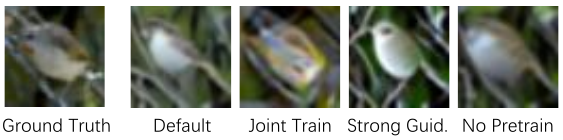}\label{subfig-ablation-attack2-egg}}
\end{tabular}
\caption{Example images of the ablation study.}
\label{fig-ablation}
\end{small}
\end{figure}

\vspace{0.5mm}\noindent
\textbf{Ablation Study of \pa.}
The three main components of \pa\ are the multi-resolution local feature extractor, the global patch smoother, and the FDN denoising module. We now evaluate the contribution of each component.
Specifically, we train three ablated variants of \pa\ on layer 17 of Gemma 3: the first variant replaces the multi-resolution extractor with a single extractor that recovers single-resolution patches; the second variant removes the global patch smoother before the FDN; and the third variant removes the FDN denoising module entirely.
The corresponding results are shown in Figure~\ref{table-ablation-attack1}, with qualitative examples provided in Figure~\ref{subfig-ablation-attack1-egg}.
From the examples in Figure~\ref{subfig-ablation-attack1-egg}, we observe the following. Compared with the original \pa, the single-resolution variant fails to accurately reproduce image colors and often loses fine-grained details. When the global patch smoother is removed, noticeable artifacts appear across patch boundaries, degrading overall visual quality. The denoising module proves to be the most critical component: the variant without FDN yields the poorest reconstruction quality, demonstrating that effective denoising is essential for producing coherent and realistic images.

\begin{table}[t]
\setlength{\tabcolsep}{3pt}
\caption{Ablation study of \pa\ on the cutting layer 17 of Gemma 3.}\label{table-ablation-attack1}
\centering
\small
\begin{tabular}{c|c|cccc}
\hline
\multirow{2}{*}{Dataset} &\multirow{2}{*}{Metric} & \multicolumn{4}{c}{Branch} \\
\cline{3-6}
  &   & Default  & $1\times$ Res. & No Smoother & No FDN \\
\hline 
\hline
\multirow{2}{*}{CIFAR10}   & $\ell_1$ ($\downarrow$) & \textbf{0.0525}& 	0.0687& 	0.0656& 	0.0894\\
                            & SSIM  ($\uparrow$) & \textbf{0.7639}	& 0.7289	& 0.7581& 	0.5655 \\    
\hline
\multirow{2}{*}{CelebA}    & $\ell_1$ ($\downarrow$) & \textbf{0.0614}& 	0.0726& 	0.0695& 	0.0865 \\
                            & SSIM  ($\uparrow$) & \textbf{0.7395}& 	0.6990	& 0.7166	& 0.6026	\\    
\hline
\multirow{2}{*}{STL10}    & $\ell_1$ ($\downarrow$) & \textbf{0.0799}	& 0.1018	& 0.0837& 	0.1355	 \\
                            & SSIM  ($\uparrow$) & \textbf{0.6639}	& 0.6187	& 0.6219& 	0.3812  \\    
\hline
\end{tabular}
\end{table}

\begin{table}[t]
\setlength{\tabcolsep}{4pt}
\caption{Ablation study of \sa\ on CIFAR10 and the cutting layer 14 of Qwen 2.5 VL.}\label{table-ablation-attack2}
\centering
\small
\begin{tabular}{c|cccc}
\hline
\multirow{2}{*}{Metric} & \multicolumn{4}{c}{Branch} \\
\cline{2-5}
    & Default  & Joint Train & Strong Guidance & No Pretrain \\
\hline 
\hline
 CSS ($\uparrow$) &\textbf{0.7842}&	0.7158&	0.7369&	0.7612\\
  $\ell_1$ ($\downarrow$) & 0.173&	0.1671&	0.2433&	0.2055\\
\hline
\end{tabular}
\end{table}

\begin{figure}[!t]
\centering
    \includegraphics[width=.6\columnwidth]{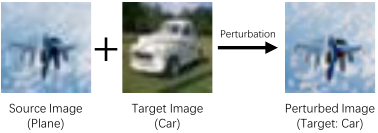}
  \caption{An adversarial example generated by AdvDiffVLM~\cite{AdvDiffVLM}.}
  \label{fig-example-adv}
\end{figure}

\vspace{0.5mm}\noindent
\textbf{Ablation Study of \sa.}
For \sa, we incorporate three main design choices to improve attack performance: the two-phase adapter and diffusion training strategy, the weak guidance strategy during diffusion training, and the diffusion pretraining stage on ImageNet.
Together, these three strategies balance reconstruction fidelity and diversity. We now evaluate the effect of each strategy separately.
For the first ablated variant, we merge the two-phase diffusion training into a single phase, meaning that the adapter and diffusion model are trained jointly.
For the second variant, we replace the weak guidance strategy with a strong guidance strategy. In this case, only a single guided diffusion model is trained, and reconstructions are generated via $\boldsymbol{s}(\x^{k}, k, \cont^{\x}) = \boldsymbol{s}_{\boldsymbol{\theta}}(\x^{k}, k, \cont^{\x})$, which resembles a simplified form of Eq.~\eqref{eq-diffusion-reconst}.
For the third variant, we remove the ImageNet pretraining stage and train the diffusion model directly from scratch.
Table~\ref{table-ablation-attack2} reports the results on layer 14 of Qwen 2.5 VL, with qualitative examples shown in Figure~\ref{subfig-ablation-attack2-egg}.
Overall, the default design of \sa\ achieves the highest semantic similarity.
While jointly training the adapter and the diffusion model yields the lowest $\ell_1$ loss, the diffusion model struggles to extract meaningful information from the adapter, resulting in semantically meaningless reconstructions.
The remaining two ablations reduce the diversity of reconstructed images: the strong guidance variant produces implausible object textures due to over-constrained generation, and the no-pretraining variant exhibits degraded background fidelity. Both effects lead to inferior semantic reconstruction compared with the full \sa\ design.

\vspace{0.5mm}\noindent
\textbf{Query Costs.}
The primary cost associated with \pa\ and \sa\ arises from the queries needed to construct the auxiliary training dataset $(\mathcal{E}^{\text{aux}}, \mathcal{D}^{\text{aux}})$.
Since each experimental dataset uses $50,000$ image-text pairs, we estimate the corresponding query expenses under three mainstream MLLM service providers: GPT-5.1~\cite{gpt5}, Gemini 2.5 Flash~\cite{gemini}, and Together.AI~\cite{togetherai}.
The estimated costs are summarized in Table~\ref{table-query-cost}.
Across all pricing models and MLLM implementations, assembling the required training dataset typically costs \textit{less than 20 USD}, highlighting the low financial barrier of the proposed attacks.

\begin{table}[t]
\setlength{\tabcolsep}{2pt}
\caption{Query costs for collecting one training dataset used in \pa\ and \sa.}\label{table-query-cost}
\centering
\small
\begin{tabular}{c|cccc}
\hline
\multirow{2}{*}{Pricing Strategy} & \multicolumn{4}{c}{MLLM Implementation} \\
\cline{2-5}
   & Gemma 3  & Phi 4 & Qwen 2.5 & Llama 4 \\
\hline 
\hline
\#Input Tokens & 13,000,000 &	13,000,000 &	400,000	& 7,400,000\\
\hline
GPT-5.1 & \$16.25	&\$16.25&	\$0.50&	\$9.25 \\    
Gemini 2.5 Flash & \$3.90 &	\$3.90 &	\$0.12 &	\$2.22 \\
Together.AI & \$0.26 &	\$0.26 &	\$0.12 &	\$1.33	\\     
\hline
\end{tabular}
\end{table}

\begin{figure}[!t]
\centering
    \includegraphics[width=.99\columnwidth]{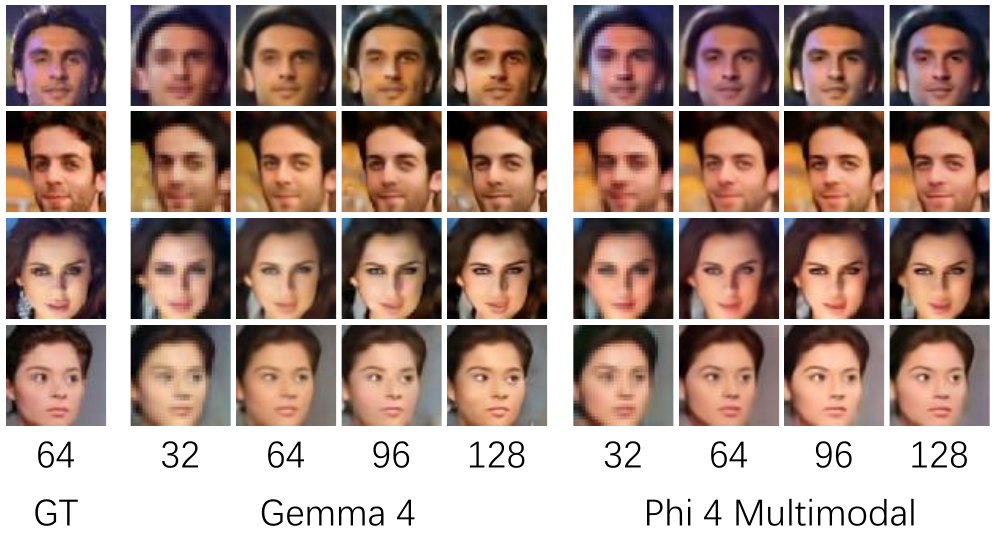}
  \caption{
      Reconstructed examples under different output resolutions. The configurations correspond to those in Table~\ref{table-diff-resolution}.
  }
  \label{fig-example-resolution}
\end{figure}

\vspace{0.5mm}\noindent
\textbf{Different Reconstruction Resolutions.}
Recall that the designs of \pa\ and \sa\ allow flexible adjustment of reconstruction resolution. We evaluate attack performance under output resolutions of $\{32 \times 32, 64 \times 64, 96 \times 96, 128 \times 128\}$. Reconstructions are performed at layer 9 of Gemma~3 and layer 8 of Phi~4~Multimodal, using CelebA as the evaluation dataset.
Table~\ref{table-diff-resolution} reports the quantitative performance of \pa, while Figure~\ref{fig-example-resolution} presents reconstructed examples. We observe that the performance of \sa\ remains largely consistent across different output resolutions in terms of the CSS metric and does not exhibit notable variation. Therefore, we focus our discussion on \pa.
Table~\ref{table-diff-resolution} and Figure~\ref{fig-example-resolution} reveal two seemingly contradictory observations. As the reconstruction resolution increases from $32$ to $128$, quantitative metrics slightly degrade (Table~\ref{table-diff-resolution}), whereas the reconstructed images exhibit improved perceptual quality to human observers (Figure~\ref{fig-example-resolution}). This discrepancy arises because higher output resolutions enable \pa\ to generate finer-grained visual details, leading to visually sharper reconstructions. However, these additional details do not necessarily correspond to the true low-level details in the original images.
This behavior is rooted in the low-level information loss introduced by MLLM processing. Specifically, the embeddings produced at a given MLLM layer may discard fine-grained visual information and instead emphasize higher-level, abstract representations. Since such information loss is irreversible, increasing the reconstruction resolution cannot recover the missing low-level details and may instead amplify mismatches between the reconstructed images and the ground truth, resulting in degraded quantitative metrics.

\begin{table}[t]
\centering
\caption{Performance of \pa\ under different reconstruction resolutions on CelebA, evaluated at layer 9 of Gemma 3 and layer 8 of Phi 4.}\label{table-diff-resolution}
\setlength{\tabcolsep}{6pt}
\centering
\small
\begin{tabular}{c|c|cccc}
\hline
\multirow{2}{*}{MLLM} &   \multirow{2}{*}{Metric }  &\multicolumn{4}{c}{Output Resolutions}  \\
\cline{3-6}
&  & 32 &  64 & 96 & 128   \\
\hline 
\hline
\multirow{3}{*}{Gemma 3}  & $\ell_1$ loss ($\downarrow$)  &	0.0412 &	0.0504 &	0.0511 &	0.0522  \\
                        & SSIM ($\uparrow$)	& 0.8823 &	0.7923 &	0.7717 &	0.7625 \\
                        & CSS ($\uparrow$)	& 0.8760 &	0.8152	&0.7631&	0.7442  \\
\hline
\multirow{3}{*}{Phi 4} & $\ell_1$ loss ($\downarrow$) &	0.0468&	0.0502&	0.0512	&0.0530  \\
                        & SSIM ($\uparrow$)	& 0.8739&	0.8112&	0.7737&	0.7380  \\
                        & CSS ($\uparrow$)	& 0.8813 & 0.8230 &	0.7719 &	0.7438 \\
\hline
\end{tabular}
\end{table}

\section{Additional Countermeasures}\label{appendix-defense}
\vspace{0.5mm}
\noindent
\textbf{Embedding Shuffle.}
Since both \pa\ and \sa\ rely on correctly matched image-embedding pairs for training, a natural defense is to disrupt this correspondence.
A centralized server could aggregate user queries into a batch, pad them to a uniform length, randomly shuffle the batch, and then feed the shuffled sequence through the inference pipeline. Under this design, an attacker would be unable to infer the correspondence between input images and embeddings based on sequence length (see Section~\ref{subsec-extract-embds}).
Nevertheless, the attacker can still identify the embedding it initiated by inserting a fragment of identifiers into the text prompt, such as a random sequence composed of special tokens (e.g., \texttt{<end\_of\_image><start\_of\_turn><start\_of\_turn>}).
While such insertion may affect generation quality, it does not interfere with the primary objective of collecting matched image-prompt pairs.
In addition, this batch-based inference defense is only practical for offline settings, where users can tolerate additional latency in LLM responses. For online inference, batching queries inevitably introduces extra delay for some users, thereby degrading service usability and response time.

\vspace{0.5mm}
\noindent
\textbf{Remove Special Tokens.}
As shown in Section~\ref{subsec-extract-embds}, both \pa\ and \sa\ rely on correctly extracting image tokens, which in turn depends on the stability of special-token embeddings, such as \texttt{<start\_of\_image>} and \texttt{<end\_of\_image>}.
Accordingly, a natural defense strategy is to remove the special tokens used in the prompt format (see Figure~\ref{fig-prompt-example}).
However, modern MLLMs universally employ special tokens as delimiters for multi-modal inputs, enabling the model to distinguish modalities and align them in a shared embedding space during training. Removing these delimiters would prevent the model from reliably separating image inputs from text inputs, thereby degrading embedding alignment and downstream performance.
Implementing such a defense would require fundamentally different training strategies and potentially new MLLM architectures, which are out of the scope of this paper.

\vspace{0.5mm}\noindent
\textbf{Differential Privacy.}
Differential privacy (DP) is a widely adopted privacy-preserving framework with strong theoretical guarantees~\cite{jiang2024calibrating-dp}. However, DP is fundamentally incompatible with the distributed MLLM inference pipeline.
DP requires that for any two neighboring datasets differing in a single sample (an image in our case), the algorithm’s outputs must be nearly indistinguishable. Applied to MLLMs, this requirement implies that the intermediate embeddings of two distinct input images $\x_1$ and $\x_2$ must be nearly identical.
Such constraints would destroy the utility of the embeddings and render the MLLM incapable of producing meaningful outputs~\cite{sl-zhundss, luo2025prompt, luo2021feature}.
Therefore, DP does not apply to this setting.

\begin{figure*}[t]
\centering
\begin{small}
\begin{tabular}{cccc}
\multicolumn{4}{c}{\hspace{0mm} \includegraphics[height=4.5mm]{figures/attack1-line-diff-layers-legend.pdf}}
\vspace{-4mm}  \\
\hspace{-5mm}
\subfloat[{Gemma 3}]{\includegraphics[width=0.25\textwidth]{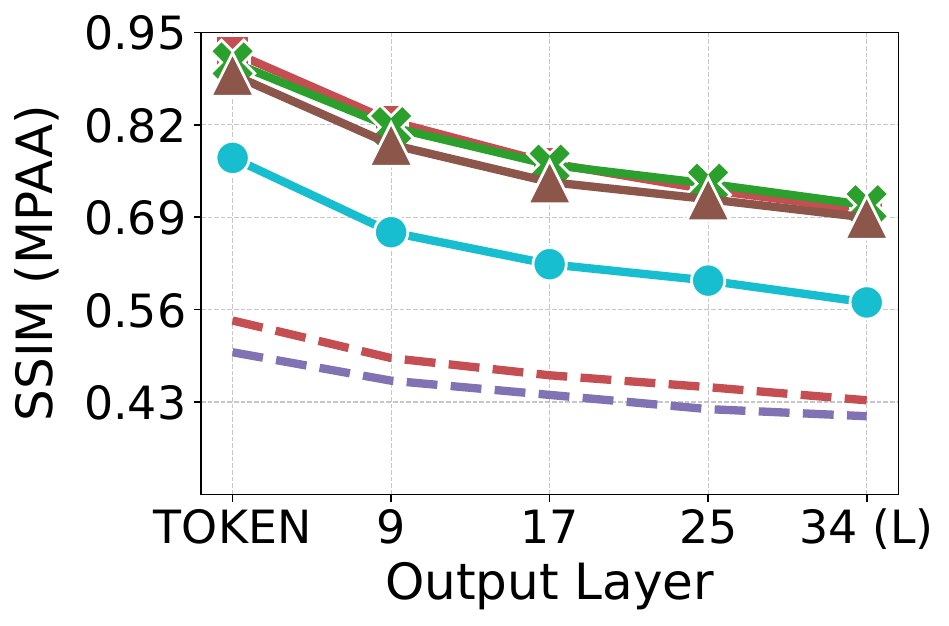}\label{}}
&
\hspace{-5mm}
\subfloat[{Phi 4 Multimodal}]{\includegraphics[width=0.25\linewidth]{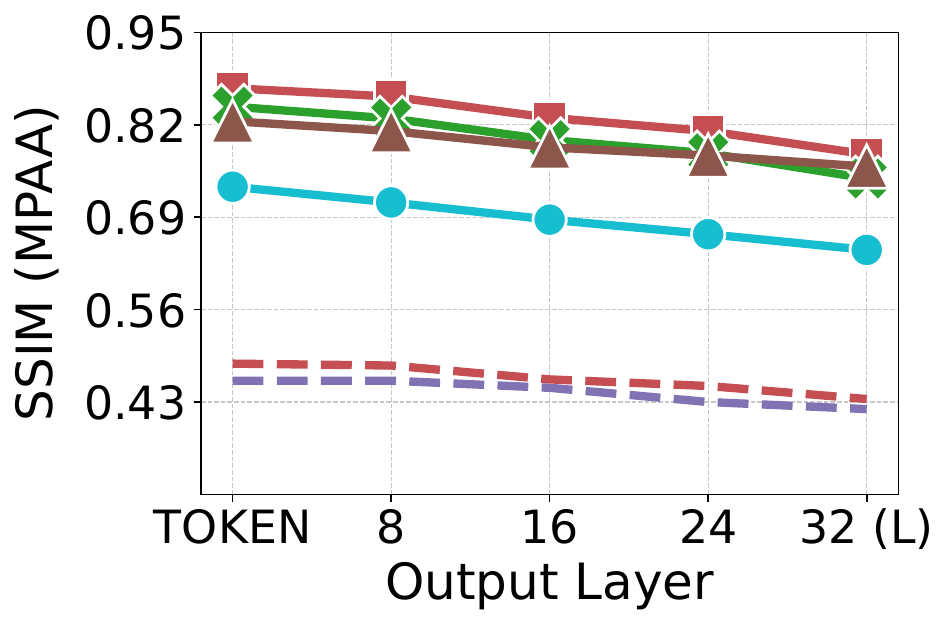}}
&
\hspace{-5mm}
\subfloat[{{Qwen 2.5 VL}}]{\includegraphics[width=0.25\linewidth]{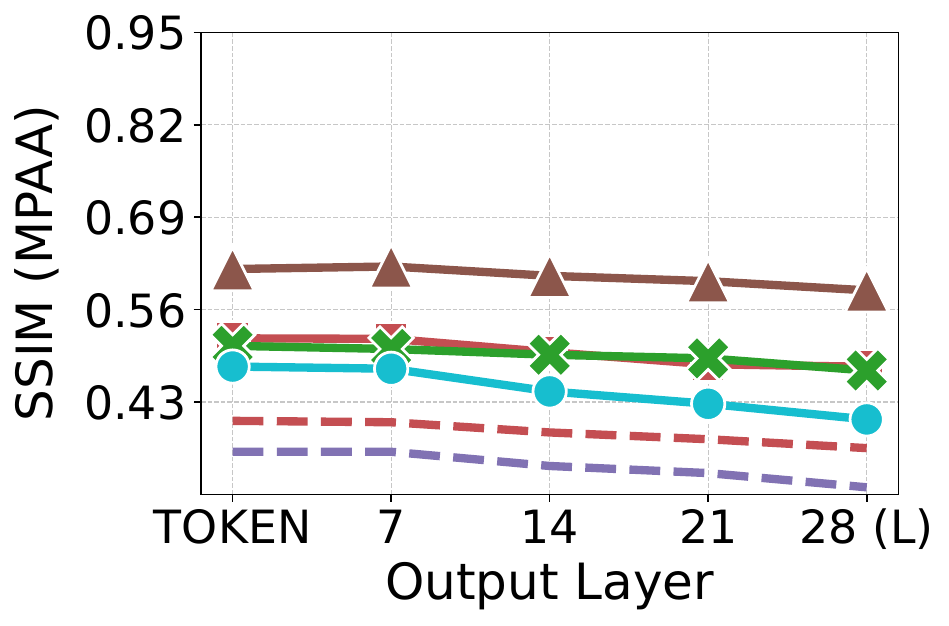}}
&
\hspace{-5mm}
\subfloat[{Llama 4 Scout}]{\includegraphics[width=0.25\textwidth]{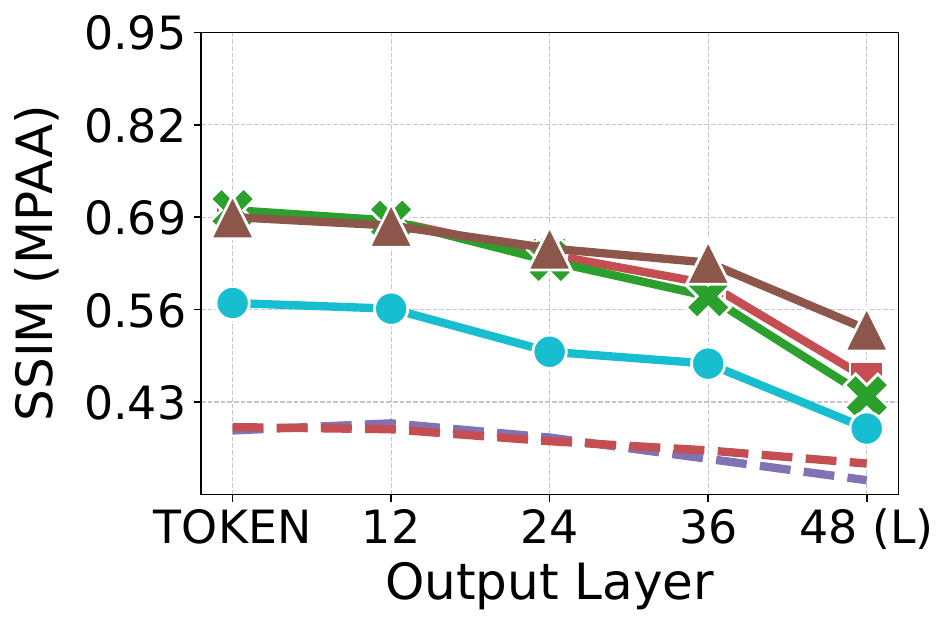}\label{}}
\end{tabular}
\caption{
The SSIM score ($\uparrow$) of \pa ((a)--(b)) across different MLLMs and datasets. It is important to note that we omit $\ell_1$ loss and SSIM results for the semantic attack (\sa), as these pixel-level metrics cannot accurately evaluate semantic reconstruction quality.
}
\label{fig-attack-loss-appendix}
\end{small}
\end{figure*}

\begin{figure*}[!th]
\centering
    \includegraphics[width=.95\textwidth]{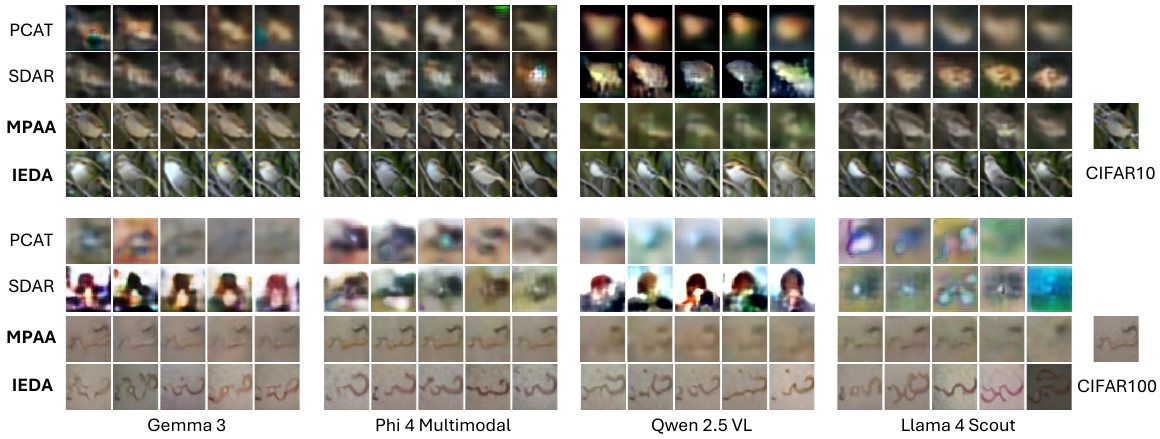}
  \caption{Example images reconstructed by different methods. In each block, the five images from left to right indicate results from the cutting layers $\{0, \lceil \frac{L}{4}\rceil, \lceil \frac{2L}{4}\rceil, \lceil \frac{3L}{4}\rceil, L\}$.}
  \label{fig-attack-examples-appendix}
\end{figure*}

%% file: tex/ethics.tex
\section*{Ethical Considerations}

This work examines privacy vulnerabilities in distributed inference frameworks for multimodal large language models (MLLMs). Given the dual-use nature of security research, we conduct a stakeholder-based ethics analysis to assess potential impacts, harms, and benefits associated with both the research process and the publication of our findings.

\noindent
\textbf{Stakeholders.}
The primary stakeholders include (1) end users whose private images may be processed by distributed MLLM inference systems, (2) developers and operators of distributed inference frameworks, (3) model providers whose architectures may be affected by privacy leakage, (4) the research community, and (5) society at large, including both defenders and potential adversaries.

\noindent
\textbf{Ethical Principles.}
Our analysis is guided by the ethical principles articulated in the Menlo Report, including \emph{Beneficence}, \emph{Respect for Persons}, \emph{Justice}, and \emph{Respect for Law and Public Interest}. In particular, we consider whether the research reduces or exacerbates privacy risks, whether it respects the rights and expectations of affected users, and whether the benefits of disclosure outweigh potential harms.

\noindent
\textbf{Potential Harms.}
The primary potential harm of this work is that insights into image reconstruction from intermediate embeddings could be misused by adversaries to infer sensitive visual information. Such misuse could result in violations of user privacy if deployed irresponsibly. However, we do not conduct experiments on real user data. All experiments are performed using publicly available models and datasets, and no personal or sensitive user information is collected or processed.

\noindent
\textbf{Mitigations.}
We take several steps to mitigate potential harms. First, we adopt a passive and black-box threat model that reflects realistic adversarial capabilities, avoiding invasive or disruptive actions. Second, we refrain from releasing turnkey attack tools that could directly enable misuse; the released code is intended for research reproducibility and requires substantial expertise and controlled conditions to apply. Third, we focus on analyzing systemic privacy risks rather than optimizing attacks for maximal exploitation. Finally, we responsibly disclosed high-level findings to the Petals developers prior to submission to support mitigation efforts.

\noindent
\textbf{Decision to Publish.}
We conclude that publishing this work is ethically justified. From a beneficence perspective, our findings expose previously unexplored privacy risks in distributed MLLM inference and provide insights that can guide the design of more privacy-resilient systems. From a respect-for-persons perspective, identifying and communicating these risks supports users’ rights to understand and protect their privacy. While acknowledging the potential for misuse, we determine that the defensive and societal benefits of transparency and early disclosure outweigh the risks, especially given the mitigations described above.

\noindent
\textbf{Human Subjects and Legal Considerations.}
This study does not involve human subjects, interaction with users, or collection of personal data, and therefore does not require IRB approval. All experiments are conducted in compliance with applicable laws and using resources permitted by their respective licenses and terms of use.


%% file: ref.bib
@inproceedings{Petals-acl,
  author       = {Alexander Borzunov and
                  Dmitry Baranchuk and
                  Tim Dettmers and
                  Maksim Riabinin and
                  Younes Belkada and
                  Artem Chumachenko and
                  Pavel Samygin and
                  Colin Raffel},
  title        = {Petals: Collaborative Inference and Fine-tuning of Large Models},
  booktitle    = {Proc. ACL},
  pages        = {558--568},
  year         = {2023},
}

@article{Petals-nips,
  title={Distributed inference and fine-tuning of large language models over the internet},
  author={Borzunov, Alexander and Ryabinin, Max and Chumachenko, Artem and Baranchuk, Dmitry and Dettmers, Tim and Belkada, Younes and Samygin, Pavel and Raffel, Colin A},
  journal={NeurIPS},
  year={2024}
}

@misc{cake,
    author = {Evilsocket},
    title = {Cake: a Rust framework for distributed inference of large models based on Candle},
    year = {2025},
    howpublished  = {\url{https://github.com/evilsocket/cake}}
}

@misc{togetherai,
    author = {Together AI},
    title = {The AI Acceleration Cloud},
    year = {2025},
    howpublished  = {\url{https://www.together.ai/}}
}

@misc{Modal,
    author = {Modal},
    title = {AI infrastructure that developers love},
    year = {2025},
    howpublished  = {\url{https://modal.com/}}
}

@misc{primeintel,
    author = {Prime Intellect},
    title = {Find Compute. Train models. Co-Own Intelligence},
    year = {2025},
    howpublished  = {\url{https://www.primeintellect.ai/}},
    note = {Online; accessed 22-August-2025}
}

@article{zhang2024edgeshard-llminference,
  title={EdgeShard: Efficient LLM Inference via Collaborative Edge Computing},
  author={Zhang, Mingjin and Cao, Jiannong and Shen, Xiaoming and Cui, Zeyang},
  journal={arXiv:2405.14371},
  year={2024}
}

@inproceedings{icml-hexgen-distributedllm,
  author       = {Youhe Jiang and
                  Ran Yan and
                  Xiaozhe Yao and
                  Yang Zhou and
                  Beidi Chen and
                  Binhang Yuan},
  title        = {HexGen: Generative Inference of Large Language Model over Heterogeneous
                  Environment},
  booktitle    = {ICML 2024},
  year         = {2024},
}

@article{LinguaLinked-distributedllm,
  author       = {Junchen Zhao and
                  Yurun Song and
                  Simeng Liu and
                  Ian G. Harris and
                  Sangeetha Abdu Jyothi},
  title        = {LinguaLinked: {A} Distributed Large Language Model Inference System
                  for Mobile Devices},
  journal      = {CoRR},
  volume       = {abs/2312.00388},
  year         = {2023},
}

@inproceedings{sigcomm-pipellm-distributedllm,
  author       = {Ruilong Ma and
                  Jingyu Wang and
                  Qi Qi and
                  Xiang Yang and
                  Haifeng Sun and
                  Zirui Zhuang and
                  Jianxin Liao},
  title        = {Poster: PipeLLM: Pipeline {LLM} Inference on Heterogeneous Devices
                  with Sequence Slicing},
  booktitle    = {Proc. ACM SIGCOMM 2023},
  pages        = {1126--1128},
  publisher    = {{ACM}},
  year         = {2023},
}

@inproceedings{qu2025prompt,
author = { Qu, Wenjie and Zhou, Yuguang and Wu, Yongji and Xiao, Tingsong and Yuan, Binhang and Li, Yiming and Zhang, Jiaheng },
booktitle = {SP 2025},
title = {{ Prompt Inversion Attack against Collaborative Inference of Large Language Models }},
year = {2025},
pages = {1602-1619},
}

@article{luo2025prompt,
  title={Prompt Inference Attack on Distributed Large Language Model Inference Frameworks},
  author={Luo, Xinjian and Yu, Ting and Xiao, Xiaokui},
  journal={arXiv:2503.09291},
  year={2025}
}

@article{qwe2.5vl,
      title={Qwen2.5-VL Technical Report}, 
      author={Shuai Bai and Keqin Chen and Xuejing Liu and Jialin Wang and Wenbin Ge and Sibo Song and Kai Dang and Peng Wang and Shijie Wang and Jun Tang and Humen Zhong and Yuanzhi Zhu and Mingkun Yang and Zhaohai Li and Jianqiang Wan and Pengfei Wang and Wei Ding and Zheren Fu and Yiheng Xu and Jiabo Ye and Xi Zhang and Tianbao Xie and Zesen Cheng and Hang Zhang and Zhibo Yang and Haiyang Xu and Junyang Lin},
      year={2025},
      journal={arXiv preprint arXiv:2502.13923},
}

@article{phi4multimodal,
  title={Phi-4-mini technical report: Compact yet powerful multimodal language models via mixture-of-loras},
  author={Abouelenin, Abdelrahman and Ashfaq, Atabak and Atkinson, Adam and Awadalla, Hany and Bach, Nguyen and Bao, Jianmin and Benhaim, Alon and Cai, Martin and Chaudhary, Vishrav and Chen, Congcong and others},
  journal={arXiv preprint arXiv:2503.01743},
  year={2025}
}

@misc{llama4,
    author = {Meta},
    title = {The Llama 4 herd: The beginning of a new era of natively multimodal AI innovation},
    year = {2025},
    howpublished  = {\url{https://ai.meta.com/blog/llama-4-multimodal-intelligence/}},
}

@article{gemma3,
      title={Gemma 3 Technical Report}, 
      author={Gemma Team and Aishwarya Kamath and Johan Ferret and others},
      year={2025},
      journal={arXiv preprint arXiv:2503.19786},
}

@inproceedings{clip,
  title={Learning transferable visual models from natural language supervision},
  author={Radford, Alec and Kim, Jong Wook and Hallacy, Chris and Ramesh, Aditya and Goh, Gabriel and Agarwal, Sandhini and Sastry, Girish and Askell, Amanda and Mishkin, Pamela and Clark, Jack and others},
  booktitle={International conference on machine learning},
  pages={8748--8763},
  year={2021},
  organization={PmLR}
}

@inproceedings{sl-tiger,
  author       = {Dario Pasquini and
                  Giuseppe Ateniese and
                  Massimo Bernaschi},
  title        = {Unleashing the Tiger: Inference Attacks on Split Learning},
  booktitle    = {ACM CCS},
  pages        = {2113--2129},
  year         = {2021},
}

@article{sl-zhundss,
  title={Passive inference attacks on split learning via adversarial regularization},
  author={Zhu, Xiaochen and Luo, Xinjian and Wu, Yuncheng and Jiang, Yangfan and Xiao, Xiaokui and Ooi, Beng Chin},
  journal={arXiv preprint arXiv:2310.10483},
  year={2023}
}

@inproceedings{sl-pcat,
  title={$\{$PCAT$\}$: Functionality and data stealing from split learning by $\{$Pseudo-Client$\}$ attack},
  author={Gao, Xinben and Zhang, Lan},
  booktitle={Proc. USENIX Security},
  pages={5271--5288},
  year={2023}
}

@article{sl,
  author       = {Otkrist Gupta and
                  Ramesh Raskar},
  title        = {Distributed learning of deep neural network over multiple agents},
  journal      = {J. Netw. Comput. Appl.},
  volume       = {116},
  pages        = {1--8},
  year         = {2018},
}

@inproceedings{luo2022fusion,
  title={A fusion-denoising attack on instahide with data augmentation},
  author={Luo, Xinjian and Xiao, Xiaokui and Wu, Yuncheng and Liu, Juncheng and Ooi, Beng Chin},
  booktitle={Proceedings of the AAAI Conference on Artificial Intelligence},
  volume={36},
  number={2},
  pages={1899--1907},
  year={2022}
}

@misc{midjourney,
    author = {Succinctly AI},
    title = {midjourney-prompts},
    year = {2025},
    howpublished  = {\url{https://huggingface.co/datasets/succinctly/midjourney-prompts}},
    note = {Online; accessed 21-October-2025}
}

@misc{gemini,
    author = {Google},
    title = {Gemini API Pricing},
    year = {2025},
    howpublished  = {\url{https://ai.google.dev/gemini-api/docs/pricing}},
    note = {Online; accessed 14-November-2025}
}

@misc{gpt5,
    author = {OpenAI},
    title = {GPT AIP Pricing},
    year = {2025},
    howpublished  = {\url{https://openai.com/api/pricing/}},
    note = {Online; accessed 14-November-2025}
}

@inproceedings{nn-memorization,
  author       = {Chiyuan Zhang and
                  Samy Bengio and
                  Moritz Hardt and
                  Benjamin Recht and
                  Oriol Vinyals},
  title        = {Understanding deep learning requires rethinking generalization},
  booktitle    = {{ICLR}},
  year         = {2017},
}

@article{shannon1959coding,
  title={Coding theorems for a discrete source with a fidelity criterion},
  author={Shannon, Claude E and others},
  journal={IRE Nat. Conv. Rec},
  volume={4},
  number={142-163},
  pages={1},
  year={1959}
}

@article{entropy-power,
  title={The convolution inequality for entropy powers},
  author={Blachman, Nelson},
  journal={IEEE Transactions on Information theory},
  volume={11},
  number={2},
  pages={267--271},
  year={2003},
  publisher={IEEE}
}

@book{signal-processing,
  title={Vector quantization and signal compression},
  author={Gersho, Allen and Gray, Robert M},
  volume={159},
  year={2012},
  publisher={Springer Science \& Business Media}
}

@inproceedings{stable-diffusion,
  title={High-resolution image synthesis with latent diffusion models},
  author={Rombach, Robin and Blattmann, Andreas and Lorenz, Dominik and Esser, Patrick and Ommer, Bj{\"o}rn},
  booktitle={Proc. IEEE/CVF CVPR},
  pages={10684--10695},
  year={2022}
}

@article{DDPM,
  title={Denoising diffusion probabilistic models},
  author={Ho, Jonathan and Jain, Ajay and Abbeel, Pieter},
  journal={NeurIPS},
  volume={33},
  pages={6840--6851},
  year={2020}
}

@inproceedings{controlnet,
  title={Adding conditional control to text-to-image diffusion models},
  author={Zhang, Lvmin and Rao, Anyi and Agrawala, Maneesh},
  booktitle={Proc. ICCV},
  pages={3836--3847},
  year={2023}
}

@inproceedings{t2iadapter,
  title={T2i-adapter: Learning adapters to dig out more controllable ability for text-to-image diffusion models},
  author={Mou, Chong and Wang, Xintao and Xie, Liangbin and Wu, Yanze and Zhang, Jian and Qi, Zhongang and Shan, Ying},
  booktitle={Proc. AAAI},
  volume={38},
  number={5},
  year={2024}
}

@inproceedings{inceptionv3,
  title={Rethinking the inception architecture for computer vision},
  author={Szegedy, Christian and Vanhoucke, Vincent and Ioffe, Sergey and Shlens, Jon and Wojna, Zbigniew},
  booktitle={IEEE CVPR},
  pages={2818--2826},
  year={2016}
}

@article{WGAN-GP,
  title={Improved training of wasserstein gans},
  author={Gulrajani, Ishaan and Ahmed, Faruk and Arjovsky, Martin and Dumoulin, Vincent and Courville, Aaron C},
  journal={Proc. NeurIPS},
  volume={30},
  year={2017}
}

@article{CFG,
  title={Classifier-free diffusion guidance},
  author={Ho, Jonathan and Salimans, Tim},
  journal={arXiv preprint arXiv:2207.12598},
  year={2022}
}

@inproceedings{imagenet,
  title={Imagenet: A large-scale hierarchical image database},
  author={Deng, Jia and Dong, Wei and Socher, Richard and Li, Li-Jia and Li, Kai and Fei-Fei, Li},
  booktitle={Proc. IEEE CVPR},
  pages={248--255},
  year={2009},
  organization={Ieee}
}

@article{CIFAR,
  title={Learning multiple layers of features from tiny images},
  author={Krizhevsky, Alex and Hinton, Geoffrey and others},
  year={2009},
  publisher={Citeseer}
}

@inproceedings{celeba,
  author    = {Ziwei Liu and
               Ping Luo and
               Xiaogang Wang and
               Xiaoou Tang},
  title     = {Deep Learning Face Attributes in the Wild},
  booktitle = {{ICCV}},
  pages     = {3730--3738},
  year      = {2015}
}

@inproceedings{STL10,
  title={An analysis of single-layer networks in unsupervised feature learning},
  author={Coates, Adam and Ng, Andrew and Lee, Honglak},
  booktitle={{AISTATS}},
  pages={215--223},
  year={2011}
}

@article{ssim,
  title={Image quality assessment: from error visibility to structural similarity},
  author={Wang, Zhou and Bovik, Alan C and Sheikh, Hamid R and Simoncelli, Eero P},
  journal={IEEE transactions on image processing},
  volume={13},
  number={4},
  pages={600--612},
  year={2004},
  publisher={IEEE}
}

@article{vepakomma2018split,
  title={Split learning for health: Distributed deep learning without sharing raw patient data},
  author={Vepakomma, Praneeth and Gupta, Otkrist and Swedish, Tristan and Raskar, Ramesh},
  journal={arXiv preprint arXiv:1812.00564},
  year={2018}
}

@inproceedings{erdougan2022unsplit,
  title={Unsplit: Data-oblivious model inversion, model stealing, and label inference attacks against split learning},
  author={Erdo{\u{g}}an, Ege and K{\"u}p{\c{c}}{\"u}, Alptekin and {\c{C}}i{\c{c}}ek, A Erc{\"u}ment},
  booktitle={Proc. WPES},
  pages={115--124},
  year={2022}
}

@article{adv-vision-hu2025transferable,
  title={Transferable adversarial attacks on black-box vision-language models},
  author={Hu, Kai and Yu, Weichen and Zhang, Li and Robey, Alexander and Zou, Andy and Xu, Chengming and Hu, Haoqi and Fredrikson, Matt},
  journal={arXiv preprint arXiv:2505.01050},
  year={2025}
}

@article{rect-vision-lei2025drag,
  title={DRAG: Data Reconstruction Attack using Guided Diffusion},
  author={Lei, Wa-Kin and Chen, Jun-Cheng and Chen, Shang-Tse},
  journal={arXiv preprint arXiv:2509.11724},
  year={2025}
}

@article{rect-vision-chen2025leakyclip,
  title={LeakyCLIP: Extracting Training Data from CLIP},
  author={Chen, Yunhao and Wang, Shujie and Wang, Xin and Ma, Xingjun},
  journal={arXiv preprint arXiv:2508.00756},
  year={2025}
}

@inproceedings{SongR20promptinference,
  author       = {Congzheng Song and
                  Ananth Raghunathan},
  title        = {Information Leakage in Embedding Models},
  booktitle    = {ACM CCS 2020},
  pages        = {377--390},
  publisher    = {{ACM}},
  year         = {2020},
}

@inproceedings{KrishnaTPPI20modelextraction2,
  author       = {Kalpesh Krishna and
                  Gaurav Singh Tomar and
                  Ankur P. Parikh and
                  Nicolas Papernot and
                  Mohit Iyyer},
  title        = {Thieves on Sesame Street! Model Extraction of BERT-based APIs},
  booktitle    = {ICLR 2020},
  year         = {2020},
}

@inproceedings{zanella2021greymodelextraction1,
  title={Grey-box extraction of natural language models},
  author={Zanella-Beguelin, Santiago and Tople, Shruti and Paverd, Andrew and K{\"o}pf, Boris},
  booktitle={ICML},
  pages={12278--12286},
  year={2021},
  organization={PMLR}
}

@inproceedings{zhang2024effective-systemprompt,
  title={Effective prompt extraction from language models},
  author={Zhang, Yiming and Carlini, Nicholas and Ippolito, Daphne},
  booktitle={First Conference on Language Modeling},
  year={2024}
}

@inproceedings{li2023sentencepromptinference2,
  author       = {Haoran Li and
                  Mingshi Xu and
                  Yangqiu Song},
  title        = {Sentence Embedding Leaks More Information than You Expect: Generative
                  Embedding Inversion Attack to Recover the Whole Sentence},
  booktitle    = {ACL 2023},
  pages        = {14022--14040},
  year         = {2023},
}

@inproceedings{MorrisKSR23promptinference,
  author       = {John X. Morris and
                  Volodymyr Kuleshov and
                  Vitaly Shmatikov and
                  Alexander M. Rush},
  title        = {Text Embeddings Reveal (Almost) As Much As Text},
  booktitle    = {Proc. EMNLP 2023},
  pages        = {12448--12460},
  year         = {2023},
}

@inproceedings{output2prompt,
  author       = {Collin Zhang and
                  John X. Morris and
                  Vitaly Shmatikov},
  title        = {Extracting Prompts by Inverting {LLM} Outputs},
  booktitle    = {Proc. EMNLP 2024},
  pages        = {14753--14777},
  year         = {2024}
}

@inproceedings{resnet,
  title={Deep residual learning for image recognition},
  author={He, Kaiming and Zhang, Xiangyu and Ren, Shaoqing and Sun, Jian},
  booktitle={Proc. IEEE CVPR},
  pages={770--778},
  year={2016}
}

@article{mllms-look,
  title={Mllms know where to look: Training-free perception of small visual details with multimodal llms},
  author={Zhang, Jiarui and Khayatkhoei, Mahyar and Chhikara, Prateek and Ilievski, Filip},
  journal={arXiv preprint arXiv:2502.17422},
  year={2025}
}

@inproceedings{chang-etal-2025-context,
    title = "Context-Aware Membership Inference Attacks against Pre-trained Large Language Models",
    author = "Chang, Hongyan  and
      Shahin Shamsabadi, Ali  and
      Katevas, Kleomenis  and
      Haddadi, Hamed  and
      Shokri, Reza",
    booktitle = "EMNLP 2025",
    month = nov,
    year = "2025",
    publisher = "Association for Computational Linguistics",
    pages = "7299--7321",
}

@article{AdvDiffVLM,
  title={Efficient Generation of Targeted and Transferable Adversarial Examples for Vision-Language Models Via Diffusion Models},
  author={Guo, Qi and Pang, Shanmin and Jia, Xiaojun and Liu, Yang and Guo, Qing},
  journal={IEEE TIFS},
  year={2024},
  publisher={IEEE}
}

@inproceedings{mp-spdz,
  author       = {Marcel Keller},
  title        = {{MP-SPDZ:} {A} Versatile Framework for Multi-Party Computation},
  booktitle    = {ACM CCS 2020},
  pages        = {1575--1590},
  publisher    = {{ACM}},
  year         = {2020},
}

@article{lu2023bumblebee,
  title={Bumblebee: Secure two-party inference framework for large transformers},
  author={Lu, Wen-jie and Huang, Zhicong and Gu, Zhen and Li, Jingyu and Liu, Jian and Hong, Cheng and Ren, Kui and Wei, Tao and Chen, WenGuang},
  journal={Cryptology ePrint Archive},
  year={2023}
}

@article{tee,
  title={Trusted execution environments: properties, applications, and challenges},
  author={Jauernig, Patrick and Sadeghi, Ahmad-Reza and Stapf, Emmanuel},
  journal={IEEE S\&P},
  year={2020},
}

@inproceedings{luo2021feature,
  title={Feature inference attack on model predictions in vertical federated learning},
  author={Luo, Xinjian and Wu, Yuncheng and Xiao, Xiaokui and Ooi, Beng Chin},
  booktitle={2021 IEEE 37th international conference on data engineering (ICDE)},
  pages={181--192},
  year={2021},
  organization={IEEE}
}

@article{jiang2024calibrating-dp,
  title={Calibrating noise for group privacy in subsampled mechanisms},
  author={Jiang, Yangfan and Luo, Xinjian and Yang, Yin and Xiao, Xiaokui},
  journal={arXiv preprint arXiv:2408.09943},
  year={2024}
}

@article{laion5b,
  title={Laion-5b: An open large-scale dataset for training next generation image-text models},
  author={Schuhmann, Christoph and Beaumont, Romain and Vencu, Richard and Gordon, Cade and Wightman, Ross and Cherti, Mehdi and Coombes, Theo and Katta, Aarush and Mullis, Clayton and Wortsman, Mitchell and others},
  journal={Proc. NeurIPS},
  volume={35},
  pages={25278--25294},
  year={2022}
}

@misc{COCO,
      title={Microsoft COCO: Common Objects in Context}, 
      author={Tsung-Yi Lin and Michael Maire and Serge Belongie and Lubomir Bourdev and Ross Girshick and James Hays and Pietro Perona and Deva Ramanan and C. Lawrence Zitnick and Piotr Dollár},
      year={2015},
      eprint={1405.0312},
      archivePrefix={arXiv},
      primaryClass={cs.CV}
}

@software{Imagenette,
    title={Imagenette: A smaller subset of 10 easily classified classes from Imagenet},
    author={Jeremy Howard},
    publisher = {GitHub},
    url = {https://github.com/fastai/imagenette}
}

@inproceedings{CC3M,
  title={Conceptual captions: A cleaned, hypernymed, image alt-text dataset for automatic image captioning},
  author={Sharma, Piyush and Ding, Nan and Goodman, Sebastian and Soricut, Radu},
  booktitle={Proc. ACL},
  pages={2556--2565},
  year={2018}
}

@misc{LLaVA,
      title={Visual Instruction Tuning}, 
      author={Liu, Haotian and Li, Chunyuan and Wu, Qingyang and Lee, Yong Jae},
      publisher={NeurIPS},
      year={2023},
}
